\documentclass[twocolumn]{aastex631}
\usepackage{CJK}
\usepackage{verbatim}
\usepackage{amsmath}
\usepackage{soul, xcolor}

\newcommand{\RNum}[1]{\uppercase\expandafter{\romannumeral #1\relax}}

\begin{document}
\begin{CJK*}{UTF8}{gbsn}

\title{The Host Galaxy (If Any) of the Little Red Dots}

\author[0009-0003-4721-177X]{Chang-Hao Chen (陈昌灏)}
\affiliation{Kavli Institute for Astronomy and Astrophysics, Peking University, Beijing 100871, China}
\affiliation{Department of Astronomy, School of Physics, Peking University, Beijing 100871, China}

\author[0000-0001-6947-5846]{Luis C. Ho}
\affiliation{Kavli Institute for Astronomy and Astrophysics, Peking University, Beijing 100871, China}
\affiliation{Department of Astronomy, School of Physics, Peking University, Beijing 100871, China}

\author[0000-0001-8496-4162]{Ruancun Li (李阮存)}
\affiliation{Kavli Institute for Astronomy and Astrophysics, Peking University, Beijing 100871, China}
\affiliation{Department of Astronomy, School of Physics, Peking University, Beijing 100871, China}

\author[0000-0001-5105-2837]{Ming-Yang Zhuang (庄明阳)}
\affiliation{Department of Astronomy, University of Illinois Urbana-Champaign, Urbana, IL 61801, USA}

\begin{abstract} 
We investigate the host galaxy properties of eight little red dots (LRDs) selected from the JWST UNCOVER survey, applying a new technique ({\tt\string GalfitS}) to simultaneously fit the morphology and spectral energy distribution using multi-band NIRCam images covering $\sim 1-4\,\mu {\rm m}$. We detect the host galaxy in only one LRD, MSAID38108 at $z = 4.96$, which has a stellar mass $\log\, (M_*/M_{\odot}) = 8.66^{+0.24}_{-0.23}$, an effective radius $R_e=0.66^{+0.08}_{-0.05}$ kpc, and a S\'ersic index $n=0.71^{+0.07}_{-0.08}$. No host emission centered on the central point source is found in the other seven LRDs. We derive stringent upper limits for the stellar mass of a hypothetical host galaxy by conducting realistic mock simulations that place high-redshift galaxy images under the LRDs.  Based on the black hole masses estimated from the broad H$\alpha$ emission line, the derived stellar mass limits are at least a factor of 10 lower than expected from the $z \approx 0$ scaling relation between black hole mass and host galaxy stellar mass. Intriguingly, four of the LRDs (50\% of the sample) show extended, off-centered emission, which is particularly prominent in the bluer bands. The asymmetric emission of two sources can be modeled as stellar emission, but the nature of the other two is unclear.
\end{abstract}

\keywords{Early universe (435); Galaxy formation (595); High-redshift galaxies (734); Active galactic nuclei (16); AGN host galaxies (2017)}

\section{Introduction} \label{sec:intro}

The emergence of the new class of objects known as the little red dots (LRDs) in the era of the James Webb Space Telescope (JWST) has significant ramifications for a variety of topics, ranging from the small-scale physics of active galactic nuclei (AGNs) to the formation mechanisms of supermassive black holes (BHs), BH-galaxy coevolution, and more. Commonly seen in the high-redshift ($z\gtrsim 4$) Universe, the LRDs share unique features that separate them from other high-redshift galaxies and AGNs. Their spectral energy distribution (SED) exhibits a distinctive ``V-shape,'' often characterized by a blue rest-frame ultraviolet (UV) continuum ($f_\lambda \propto \lambda^{\beta}$, with $\beta_{\rm UV}\lesssim -0.4$) superposed on top of a red rest-frame optical continuum ($\beta_{\rm opt}>0$) \citep[][]{Kocevski2023, 2024arXiv240610341A, 2024arXiv240403576K, 2024ApJ...968...38K, 2023arXiv230607320L, 2024ApJ...963..129M}. LRDs are also distinguished by extremely compact morphology, with an estimated median effective radius of $R_e \approx 100$ pc \citep[][]{2024arXiv240610341A, 2024ApJ...968...38K, 2023arXiv230607320L}, or less \citep{2024Natur.628...57F}. The number density of LRDs drops sharply below $z\approx 4$, with few counterparts that have similar SED and morphology seen in the local Universe \citep{2024arXiv240403576K}. While being $\sim 100$ times more abundant compared to UV-selected quasars at similar redshift, their proportion of the total galaxy population is generally marginal ($\sim 1\%$), although the contribution can be significant ($\sim 10\%$) at the bright end ($M_{\rm UV}\approx -22$ mag) at $z\approx 7$ \citep[][]{2024arXiv240610341A, 2024arXiv240403576K, 2024ApJ...968...38K}. 

As a consequence of their compactness and red optical color, LRDs were thought to be candidates of obscured AGNs upon their initial discovery. Indeed, according to follow-up spectroscopic observations, $\sim 60\% - 80\%$ of the photometrically selected LRDs are confirmed to be at high redshifts and have broad H$\alpha$ or H$\beta$ emission lines with full width at half-maximum (FWHM) velocities $\gtrsim 1000$ $\rm km~s^{-1}$ \citep{2024ApJ...964...39G}. Such large line widths conform to the canonical definition of broad-line (type~1) AGNs \citep[e.g.,][]{1977ARA&A..15...69W, 2005AJ....129.1783H}. The broad-line components exclusively detected in permitted lines indicate not only extremely high velocities but also high densities typically associated with the central regions in the vicinity of supermassive BHs (\citealt{1989ApJ...347..640R, 1995ApJ...455L.119B, Ho2008, Kokubo2024}). This is, thus far, the most compelling evidence linking LRDs to AGNs. A significant contribution of the accretion disk is implicated from modeling of the SED, which indicates that the red slope of the optical continuum is better fit with obscured emission from an AGN than from star formation, especially when flux upper limits from ALMA millimeter observations are available \citep{2023arXiv230607320L}. Moreover, the observed luminosity of the H$\alpha$ line tracks the 5100\,\AA\ continuum as in typical type~1 AGNs \citep{2005ApJ...630..122G}, suggesting that the optical continuum is likely to be dominated by emission from an AGN accretion disk \citep{2024ApJ...964...39G}.

Despite the presence of broad emission lines, the multi-wavelength properties of LRDs, from the X-rays to the mid-infrared, deviate from those of conventional type~1 AGNs, which may shed light on nuclear structures and dust properties that are potentially different from those in the local Universe. For instance, the rest-frame near-infrared continuum of LRDs flattens at $\sim 1-2\, \mu$m \citep[][]{2024ApJ...968....4P,2024arXiv240302304W, 2024ApJ...968...34W}, inconsistent with typical AGN spectra that continue to rise because of the hot dust emission from the torus. Efforts to explain the low dust temperatures ($\sim 100$ K; \citealt{2024arXiv240705094C}) invoke a clumpy torus model with flexible geometry and clump size \citep{2024arXiv240302304W} or a grey extinction curve due to the absence of small-size dust grains \citep{2023arXiv231203065K}, coupled with a more spatially extended distribution of dust \citep{2024arXiv240710760L}. Unlike local type~1 AGNs, most LRDs are undetected in the X-rays \citep[][]{2024ApJ...969L..18A,2024arXiv240413290Y}. Although a Compton-thick environment \citep{2024arXiv240500504M} may be the culprit, the X-ray weakness of LRDs may be an intrinsic signature of highly accreting BHs \citep{2024arXiv240715915P, Madau2024}, a phenomenon already familiar in the context of nearby AGNs \citep[e.g.,][]{2012ApJ...761...73D, Wang2014}. The tentative detection of optical flux variations in at least some LRDs \citep{Zhang2024} partly alleviates earlier concerns of their apparently lack of variability \citep{Hayes2024, Kokubo2024}.

In the local Universe, tight correlations have been found between the mass of the supermassive BH and its host galaxy properties, including bulge stellar mass and velocity dispersion \citep{Magorrian1998, 2000ApJ...539L...9F, 2000ApJ...539L..13G, 2013ARA&A..51..511K}. Depending on galaxy type and galaxy mass, nearby galaxies have a characteristic BH-to-stellar mass ratio of $M_{\rm BH}/M_* \approx (0.9-6.2) \times 10^{-3}$ \citep{2020ARA&A..58..257G}. These BH-host scaling relations, when extended to low-redshift AGNs and quasars, show no significance departures from the relations established for inactive galaxies (e.g., \citealt{Greene2006, Bennert2011, 2023NatAs...7.1376Z, Molina2024}). The situation at higher redshift, on the other hand, has become increasingly complicated and contentious. While evidence has mounted for some time that high-redshift quasars appear to have systematically overmassive BHs relative to the BH-galaxy scaling relations established at low redshifts (e.g., \citealt{Walter2004, Peng2006, 2013ARA&A..51..511K, Pensabene2020}), there have been lingering concerns of whether observed differences are real or simply a consequence of selection effects (e.g., \citealt{Schulze2011}). The advent of JWST has only intensified the debate on this increasingly active field. The majority of the AGN populations discovered at $z\gtrsim 3$ so far have BHs $\sim 10 - 100$ times overmassive compare to their local counterparts \citep{2023arXiv230801230M, 2023ApJ...957L...3P}, ranging from quasars with $M_{\rm BH} \approx 10^9 \,M_{\odot}$ \citep{2024ApJ...964...90S, 2024ApJ...966..176Y} to less massive AGNs---among them LRDs---with $M_{\rm BH} \approx 10^6-10^8\, M_{\odot}$ \citep{2023ApJ...955L..24G, 2023ApJ...959...39H, Kocevski2023, 2023ApJ...953L..29L, 2023arXiv230801230M, 2024Natur.628...57F}. The prevalence of overmassive BHs at high redshifts, if not an artifact of sample selection effects \citep{2024arXiv240300074L}, suggests that the early growth of seed BHs predates or at least outpaces their host galaxy mass assembly \citep{Inayoshi2022a, Scoggins2024}. 

LRDs constitute a significant fraction of high-redshift broad-line emitters. According to \citet{2023ApJ...959...39H} and \citet{2023arXiv230801230M}, $\sim 20\%$ of high-redshift broad-line AGNs are LRDs. How they fit into the BH-galaxy coevolution scenario remains to be seen. Critically lacking is the definitive detection and accurate measurements of basic physical properties of the host galaxies of LRDs. To date, efforts to constrain the hosts of LRDs through SED and spectral decomposition suffer from the degeneracy between galaxy and AGN models \citep[e.g.,][]{2023arXiv230801230M, 2024ApJ...964...39G, 2023arXiv230607320L, 2024ApJ...969L..13W}. Even the few with a Balmer break detected in their spectra, supposing that it arises from stars and not from dense gas \citep{2024arXiv240907805I}, can have derived stellar masses that vary by 2 orders of magnitudes depending on the assumed contribution of the AGN \citep{2024arXiv240807745B, 2024arXiv240720320K, 2024ApJ...969L..13W}. This large uncertainty in the stellar mass of the host galaxy obviously strongly impacts any consideration of the high-redshift BH-host mass scaling relation.

The severe model degeneracy in spectral and SED modeling can be alleviated partly by taking the spatial structure of the source into account. The AGN emission arises from an unresolved source originating from the center of the host, while the stellar continuum should be, to some degree, more extended. Although the LRDs are, by selection, compact and unresolved sources in the reddest Near Infrared Camera \citep[NIRCam;][]{2023PASP..135b8001R} F444W filter, some are known to have resolved strutures in the bluer filters that have higher spatial resolution. For instance, \citet{2023arXiv231203065K} performed simultaneous multi-band image analysis of an LRD at $z\approx 4.5$, and found that it is best fit with a point source superposed on an extended \cite{Sersic1968} profile. The effective radius of the extended component increased from $R_e \lesssim 170$ pc in F444W to $R_e = 450$ pc in F115W. Such wavelength-dependent morphological transformation was interpreted as a red compact source residing in a blue star-forming galaxy. However, this scenario does not necessarily apply to all LRDs. \citet{2023ApJ...959...39H} performed AGN-host decomposition of LRDs using HST and JWST images but failed to detect any extended emission associated with the host, leading the authors to conclude that their sources are at least 2 orders of magnitude overmassive relative to the local BH-galaxy scaling relation. An even more extreme case was reported by \citet{2024Natur.628...57F}, who analyzed an LRD triply lensed by the foreground galaxy cluster Abell~2744. The LRD remains unresolved in all seven NIRCam bands (from F115W to F444W) even after the strong lensing magnification, with an effective radius upper limit of $R_{e}\lesssim 30$ pc. The upper limit on stellar mass, which was derived from the surface brightness profile by setting the stellar surface density equal to that of the densest star clusters or elliptical galaxy progenitors in the local Universe, also shows an exceptionally high BH-galaxy mass ratio of $M_{\rm BH}/M_{*}\gtrsim 0.03$. 

To accurately measure host galaxy stellar mass for the LRDs, SED analysis is needed alongside morphological fitting. For LRDs that are resolved in the rest-frame UV, the SED of the extended component can be extracted and analyzed separately without contamination from the nucleus. Even for those that remain unresolved in all bands, the apparent absence of the host galaxy component can be a result of the host being too faint or too compact to be distinguishable from the dominant central nucleus. In this case, more accurate stellar mass limits can be obtained through careful mock simulations. In this paper, we aim to disentangle possible extended starlight from the central point source of eight spectroscopically confirmed LRDs, six of which have robust detection of broad emission lines, and hence BH mass estimates, in order to better understand the nature of the host galaxies of the LRDs. Unlike previous works, we simultaneously fit the morphology in all available NIRCam bands while constraining the multi-band fluxes of each component using SED models. 

Section~\ref{data_sample} presents the sample selection and imaging data used. Section~3 describes our point-spread function (PSF) construction technique and the technical details of our image decomposition method.  Our main results are presented in Section~\ref{results}, which include a
strategy to estimate the upper limits of the host galaxy, and an analysis of the off-centered extended emission detected in half of the sample. The implications of our findings are discussed in Section~\ref{discussion}, before concluding in Section~\ref{summary}. We assume a flat cosmology with $\Omega_m=0.3111$, $\Omega_{\Lambda}=0.6889$, and Hubble constant $H_0=67.66$ km ${\rm s}^{-1}$ ${\rm Mpc}^{-1}$ \citep{2020A&A...641A...6P}.

\begin{figure*}[ht!]
\includegraphics[width=\textwidth]{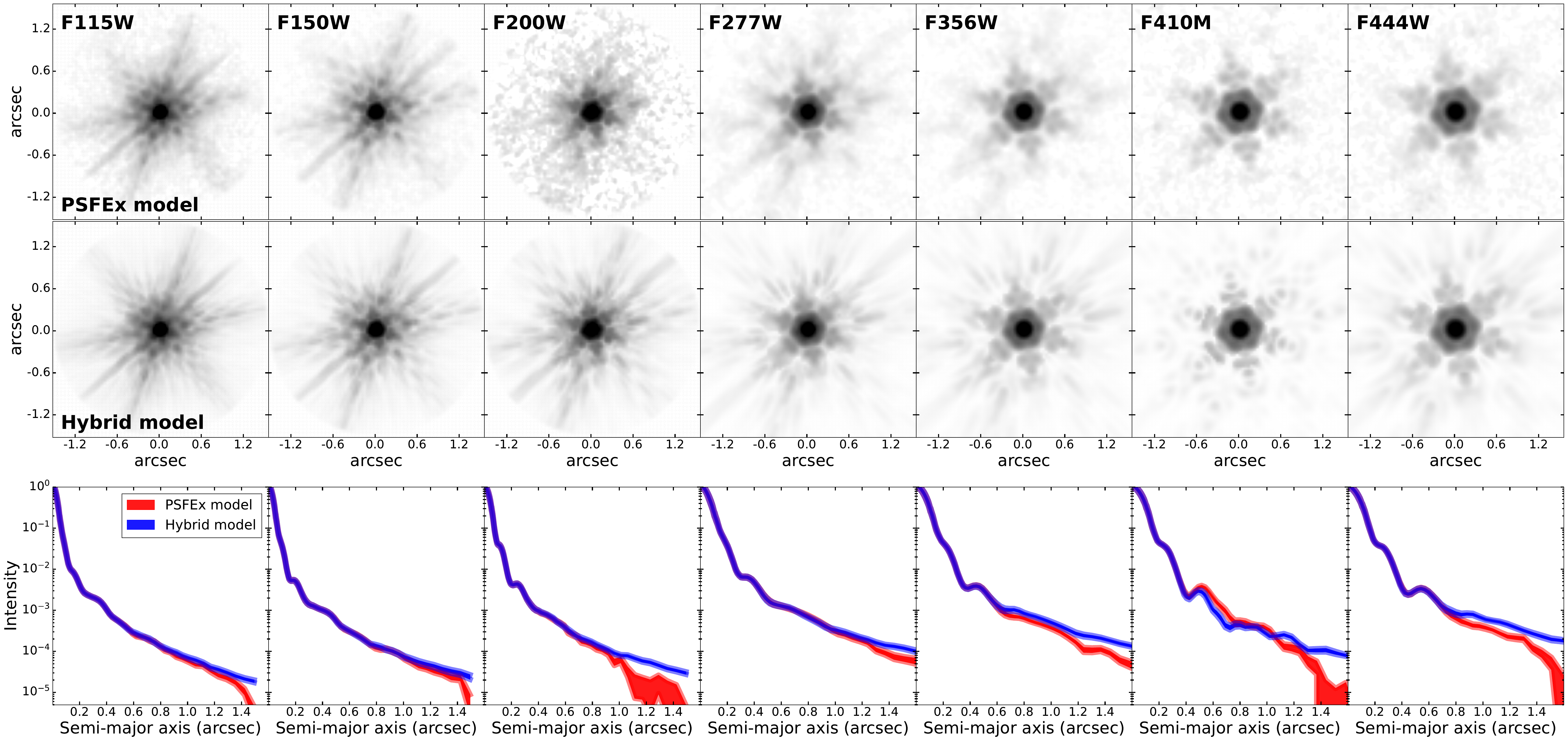}
\caption{PSF models for the seven NIRCam bands, constructed empirically using {\tt\string PSFEx} (top) and using the hybrid technique described in Section~\ref{hyb_psf} (middle). The bottom row shows the 1D surface brightness profiles for the {\tt\string PSFEx} (red) and hybrid (blue) PSF models.
\label{psf_demo}}
\end{figure*}

\section{Data and Sample} \label{data_sample}

The data used in this work were taken as part of the JWST Cycle~1 Treasury program Ultradeep NIRSpec and NIRCam ObserVations before the Epoch of Reionization (UNCOVER; \citealp{2022arXiv221204026B}), which includes ultra-deep ($\rm F150W \approx 29-30$ mag) imaging observation of $\sim$45 ${\rm arcmin}^2$ around the galaxy cluster Abell~2744 at $z = 0.308$. We utilize the multi-band mosaic images from UNCOVER data release 2 (DR2) reduced by \citet{2022arXiv221204026B} and \citet{2024arXiv240413132S}, in which they aligned the images to stars from the Gaia DR3 catalog (\citealp{2023A&A...674A...1G}), co-added all the flux-calibrated Stage 2 images, and subtracted large-scale background using the {\tt\string Grizli} pipeline \citep{2019ascl.soft05001B, 2022zndo...6672538B}. Images in the short-wavelength (SW: F090W, F115W, F150W, and F200W) and long-wavelength (LW: F277W, F335M, F356W, F410M, and F444W) bands are drizzled to a pixel scale of 0\farcs02 and 0\farcs04, respectively. Their final data products include mosaics from nine NIRCam filters.

\cite{2023arXiv230607320L} selected 40 LRDs by applying color criteria that select a distinctive ``V-shaped" SED, as well a compactness criterion measured in the F444W-band image, $f_{{\rm F444W}}(0\farcs4)/f_{{\rm F444W}}(0\farcs2)<1.7$. They then performed a joint PSF$+$S\'ersic fit of the F444W-band images using {\tt\string Galfit} \citep{Peng2002,Peng2010} and identified 26 point source-dominated targets in which more than 50\% of the flux comes from an unresolved point source. Finally, out of these point source-dominated targets, an ``SED-selected" sample consisting of 17 LRDs was selected by keeping only sources whose SED fitting shows significant improvement if an AGN component is included. \cite{2024ApJ...964...39G} carried out NIRSpec/PRISM follow-up observations with the Micro-Shutter Assembly (MSA) that prioritized the 17 SED-selected LRDs. Among the targets observed in \cite{2024ApJ...964...39G} with spectroscopic follow-up, three were found to be brown dwarfs and three others turned out to be triply lensed images of the same source, which leaves 12 sources of extragalactic origin. \cite{2024ApJ...964...39G} identified nine robust broad-line AGNs from their emission-line analysis, by requiring that the $\chi^2$ statistic \citep{Bevington 1969} of the fit improved by more than $3\,\sigma$ after including a broad component wih $\rm FWHM > 2000$ $\rm km~s^{-1}$ to the H$\alpha$ line detected at greater than $5\,\sigma$ significance.

The targets used in this work are drawn from the 12 extragalactic sources observed in \cite{2024ApJ...964...39G}. We first require that our targets to have lensing magnification $\mu<2$ in order to minimize morphological distortion. Our final sample consists of eight sources, including six robust broad-line AGNs and two unconfirmed AGN candidates: MSAID2008 does not show significant $\chi^2$ improvement after including a broad-line component, while MSAID10686 has $< 5\, \sigma$ broad-line detection. Nevertheless, we still include these two AGN candidates in the following analysis. Images in seven NIRCam bands (F115W, F150W, F200W, F277W, F356W, F410M, and F444W) are available for all the targets in our sample, except for MSAID4286, which does not have coverage in F410M mosaics.

\section{Image Decomposition}\label{image_decomposition}

\subsection{PSF Construction} \label{PSF}

Accurate PSFs are critical for morphological analysis, especially when all the LRDs in our sample are, by selection, barely resolved in F444W. Methods for PSF modeling adopted in previous works largely can be classified into two categories, namely theoretical modeling or empirical modeling. Several works, particularly those based upon the earliest NIRCam data releases \citep[e.g.,][]{2023ApJ...946L..15K, 2023ApJ...948L..18N, 2023ApJ...950....7S, 2023arXiv231203065K}, simulated theoretical PSFs using the Python package {\tt\string WebbPSF}\footnote{\url{https://webbpsf.readthedocs.io/en/latest/intro.html}} \citep{2012SPIE.8442E..3DP, 2014SPIE.9143E..3XP, 2015ascl.soft04007P}. However, since the {\tt\string WebbPSF} pipeline does not include the drizzling process, it generally produces theoretical PSFs with narrower FWHM compare to star cutouts from drizzled mosaics \citep{2022ApJ...939L..28D, 2023ApJ...951...72O}. This problem can be mitigated effectively using empirical PSF models constructed from a library of field star images. According to \citet{2000PASP..112.1360A}, each star image can be seen as a sub-pixel sampling of the PSF. By extracting information from a group of field star images that sufficiently sample the central sub-pixel position, empirical PSF can be modeled on a finer sub-pixel grid. 

\subsubsection{Empirical PSF Model} \label{emp_psf}

Several algorithms can construct empirical PSF models from star images. First is image stacking, which co-adds together field star images to obtain PSF models. \citet{2000PASP..112.1360A} proposed an alternative method that iteratively solves for the sub-pixel center of each field star using the stacked star image as initial guess to the PSF model. In each iteration, the median fitting residual is added back to the current PSF model for correction, and the process terminates once both the star center positions and the PSF model solution converge. The pixel-based algorithm in the {\tt\string PSFEx} code \citep{2013ascl.soft01001B} uses a similar principle as \citet{2000PASP..112.1360A}, but instead of directly solving the pixel value for the PSF model, it fits the difference between the median stacked star images and the real PSF model. Comparing these three methods, \citet{2024ApJ...962..139Z} conclude that PSF models created by {\tt\string PSFEx} have the highest resemblance to field star images, both in terms of the recovered magnitude and fitting $\chi^2$. We therefore generate our empirical PSF model from field star images using this method. 

We first use {\tt\string SExtractor} \citep{1996A&AS..117..393B} to identify point-like sources within the UNCOVER seven-band mosaics following the parameter set-up described in \citet{2024ApJ...962..139Z}. Specifically, we select candidate stars that are of high signal-to-noise ratio ({\tt\string SNR\_WIN} $>100$), unblended ({\tt\string FLAGS} $<2$), regular in shape ({\tt\string ELONGATION} $<1.5$), point-like ({\tt\string CLASS\_STAR} $>0.8$), and not contaminated by bad pixels ({\tt\string IMAFLAGS\_ISO} $=0$). True field stars are then selected by fitting their SEDs to a library of stellar spectra, following \citet{2024ApJ...960..104S}. The SED of each point source comprises 10 photometric points from the catalog by \citet{2024arXiv240413132S}, three (F115W, F150W, and F200W) from NIRCam and seven (F435W, F606W, F814W, F105W, F125W, F140W, and F160W) from the Wide Field Camera 3 (WFC3; \citealt{Dressel2019}) on the Hubble Space Telescope. The observed SEDs of the point sources are then matched to stellar spectral templates from the ESO Library of Stellar Spectrum\footnote{\url{https://www.eso.org/sci/facilities/paranal/decommissioned/isaac/tools/lib.html}} using least-squares fitting. Similar to \citet{2024ApJ...960..104S}, we find that most of our selected stars are G-type or K-type giants. We also explore a different star selection method from \citet{2015MNRAS.448.1305K}, who adopt a color selection using the WISE W1 and 2MASS $J$ bands: ${\rm W1}-J>-1.7$ mag. As the effective wavelength of the W1 and $J$ filters are broadly consistent with NIRCam F356W and F115W, respectively, we adopt $\rm F356W - F115W >-1.7$ mag. We find excellent agreement between the star sample selected from SED fitting and from color selection. Compare to SED fitting, color selection requires less computational cost and only uses NIRCam photometry, offering an alternative to SED selection in future JWST deep fields when photometric data bluer than $1\,\mu$m might not be available. 

We conducted several tests to verify that the selected field stars provide a reliable representation of the intrinsic PSF. First, the Hawaii-2RG HgCdTe detectors used in NIRCam are known to exhibit the brighter-fatter effect \citep[e.g.,][]{Plazas2018}, wherein the PSF FWHM increases with the brightness of the point source. However, for our star sample, no significant correlation was found between FWHM and signal-to-noise ratio (SNR), possibly because of the limited dynamical range of the SNR.  In addition, the SEDs of LRDs differ from those of the stars used for the PSF modeling. To evaluate the impact of PSF spectral variation on our analysis, we utilized {\tt\string WebbPSF} to simulate two PSF models: one based on the spectrum of a K5 III star (star PSF model) and the other derived from the LRD continuum, interpolated from the broad-band photometry of MSAID10686 (LRD PSF model). The LRD PSF model was scaled to a magnitude of 26, typical for our LRD sample, and added to an empty sky region to create a mock LRD point source. When subtracting the star PSF model from the mock LRD point source image, the residuals appeared smooth and lacked discernible patterns. We therefore conclude that neither the brighter-fatter effect nor the spectral difference have a substantial impact on the construction of our PSF model.

We create the UNCOVER image mosaics by co-adding observations with different telescope on-sky position angles. Since all datasets are aligned first before stacking, the shape of the PSF in certain overlapping regions will be altered because the point source images are also rotated and stacked. Among the eight LRDs initially selected, four (MSAID2008, 10686, 20466, and 41225) are stacked from datasets with the same on-sky position angle (PA $= 41^{\circ}$), for which we identified 11 approporiate stars for PSF modeling. Two targets are stacked from datasets with two different position angles (PA $= 41^{\circ}$ and $44^{\circ}$ for MSAID45924; PA $= 59^{\circ}$ and $29^{\circ}$ for MSAID4286). Two other targets are stacked from datasets with three different position angles (PA $= 59^{\circ}$, $29^{\circ}$, and $41^{\circ}$ for MSAID13821; PA $= 43^{\circ}$, $41^{\circ}$, and $251^{\circ}$ for MSAID38108). Among the targets stacked from multiple position angles, we can only identify six stars using the same datasets as MSAID45924, and there are no suitable stars for the other three targets (MSAID4286, 13821, and 38108). Empirical PSF models for the five LRDs with suitable stars are constructed from their corresponding star images using {\tt\string PSFEx} following the parameter configuration described in \citet{2024ApJ...962..139Z}. For the remaining three LRDs, we take the empirical PSF created for targets stacked with a single position angle, and in turn rotate them to match the multiple position angles used to create mosaics for them. Then, the rotated empirical PSF models are stacked with their centers aligned in order to obtain the final empirical PSFs for these three LRDs with no suitable stars. Rotation of the empirical PSFs is done on a twice super-sampled pixel grid in order to minimize distortions from the rotation algorithm. Due to the small number of stars identified, we do not consider PSF spatial variation across the detector, and only create a global PSF model for each LRD.

\begin{deluxetable*}{ccccccccc}  
\label{results_table}
\tablecaption{Fitting Results of LRDs} 
\tablewidth{0pt}
\tablehead{MSAID & $z$ & $\mu$ & Model & Prior & BIC &  log~($M_{\rm *}/M_{\odot}$) &  log~($M_{\rm BH}/M_{\odot}$) \\ 
		(1) & (2) & (3) & (4) & (5) & (6) & (7) & (8) &}
\startdata    \label{table:results}
     &&& S\'ersic+PS & MSR & 5086.19 &  & \\
2008 & 6.74 & 1.69 & S\'ersic+PS & no prior & 5086.20 &  $<8.68$ & \nodata \\
     &&& PS & no prior & 4995.68 & & & \\ 
 \hline
     &&& S\'ersic+PS & MSR & 5239.90 &   & \\ 
4286 & 5.84 & 1.62 & S\'ersic+PS & no prior & 5236.31 &  $<8.43$ &  $8.00\pm 0.30$\\ 
     &&& PS & no prior & 5150.10 & & & \\ 
 \hline
     &&& S\'ersic+PS & MSR & 5935.94 &  & & \\ 
10686 &5.05 & 1.44 & S\'ersic+PS & no prior & 5966.00 & $<8.32$ & \nodata \\ 
     &&& PS & no prior & 5891.44 & && \\ 
 \hline
     &&& S\'ersic+PS & MSR & 5598.03 &  &\\ 
13821 & 6.34 & 1.59 & S\'ersic+PS & no prior & 5594.74 &  $<9.46$ & $8.10\pm 0.20$\\ 
     &&& PS & no prior & 5561.42 & & & \\ 
 \hline
     &&& S\'ersic+PS & MSR & 4601.82 &   &  \\ 
20466 &8.50 & 1.33 & S\'ersic+PS & no prior & 4601.82 & $<8.70$ & $8.84\pm 0.42$ \\ 
     &&& PS & no prior & 4511.35 &  & & \\ 
 \hline
     &&& S\'ersic+PS & MSR & 3937.91 &  &  \\ 
38108 &4.96 & 1.59 & S\'ersic+PS & no prior & 3956.39 & $8.66^{+0.24}_{-0.23}$ & $8.40\pm 0.50$\\ 
     &&& PS & no prior & 5776.27 & & & \\ 
 \hline
     &&& S\'ersic+PS & MSR & 6892.03 &   &\\ 
41225 & 6.76 & 1.50 & S\'ersic+PS & no prior & 6884.67 &  $<9.48$ & $7.70\pm 0.40$\\ 
     &&& PS & no prior & 6801.44 & & & \\ 
 \hline
     &&& S\'ersic+PS & MSR & 103455.34 &  & \\ 
45924 &4.46 & 1.59 & S\'ersic+PS & no prior & 116386.90 &  $<9.60$ &  $8.90\pm 0.10$\\ 
     &&& PS & no prior & 63498.69 & & & \\
 \hline
 \enddata
 
\tablecomments{Col. (1): Object ID in MSA configuration \citep{2024ApJ...964...39G}. Col. (2): Spectroscopic redshift. Col. (3): Total lensing magnification, adopted from \citet{2023MNRAS.523.4568F}. Col. (4): Morphological model adopted for the fit, with ``PS'' denoting point source. Col. (5): Astrophysical prior adopted. Col. (6): Bayesian information criterion value for each fit. Col. (7): Host galaxy stellar mass derived from our analysis. Col. (8): Single-epoch BH mass \citep{2024ApJ...964...39G}.}
\end{deluxetable*}

\subsubsection{Hybrid PSF Model} \label{hyb_psf}

Stellar SEDs peak at around $1-1.5 \,\mu$m and then decline with increasing wavelength, making NIRCam star images gradually fainter in redder bands. Therefore, the signal-to-noise ratio of the empirical PSF models varies significantly, with lower signal-to-noise ratio in long-wavelength filters. It is crucial to unify the quality of the PSF across different bands in order to perform accurate morphological analysis through simultaneous, multi-band fitting (Section~\ref{gs_fitting}). We consider hybrid PSF modeling with this goal in mind.

Theoretical PSF models for NIRCam generated from {\tt\string WebbPSF} are narrower compare to actual star images and cannot describe well the central profile of a point source \citep{2022ApJ...939L..28D, 2023ApJ...951...72O, 2024ApJ...960..104S, 2024ApJ...962..139Z}. Nevertheless, theoretical modeling offers the only practical means of securing useful information for the extended wings of the PSF that are otherwise difficult to recover under the noise in actual star images. Despite their poor performance in the core, simulated PSF models for space telescopes can produce relatively well the outer regions of the PSF, from the first Airy ring to the diffraction spikes \citep{2012ApJS..203...24V}. To ensure that the theoretical PSF models from {\tt\string WebbPSF} show the same diffraction spikes as real observed stars, we use the {\tt\string WebbPSF} model to fit the F115W image of the brightest star in our sample, which shows the most prominent outer diffraction spikes. We find that the outer wings, typically between $0\farcs5$ and $1\arcsec$, can be well represented by the model. The basic idea of hybrid PSF modeling is to combine the central part of an empirical PSF model with the outer part of a theoretical PSF model so that an accurate depiction of the entire PSF, from its core to its wings, can be obtained. A similar strategy was employed by \citet{2012ApJS..203...24V} in their treatment of WFC3 data, in which the central pixels (within a radius of three pixels) of the simulated PSF was replaced with the same region of the median-stacked star image. The resulting hybrid PSF model shows better agreement with field star images.

In this work, we adopt the same principle but with a slightly different approch. Specifically, for each star, we calculate the local background standard deviation from nearby empty regions. The background level of the empirical PSF model is taken as the median value of these local background standard deviations from stars used to construct the empirical PSF. For each pixel, if its value is below the background level of the empirical PSF, it is replaced by the value of the same pixel on the theoretical PSF model, which has been rotated and stacked in advance to match the position angles of the drizzled datasets. To redistribute the light fraction between the central empirical part and the outer theoretical part, we require the one-dimensional (1D) surface brightness profile of the hybrid PSF model to be continuous. Defining $r_{c}$ as the radius where the average surface brightness of the empirical PSF equals to the fluctuation level of the background surface brightness,

\begin{equation}
\mu_{\rm emp}(r_{c})=\sigma_{\rm sky},
\end{equation}

\noindent
we calculate the scale factor $f$ as the ratio between the surface brightness of the empirical and theoretical PSF at $r_{c}$:

\begin{equation}
f=\mu_{\rm emp}(r_{c})/\mu_{t}(r_{c}).
\end{equation}

\noindent
Then, for all the pixels replaced by the theoretical PSF value, we multiply their replaced pixel value with the factor $f$, so that the 1D surface brightness profile of the hybrid PSF becomes continuous after connecting the central and outer parts. The hybrid PSF model is then normalized so that the sum of all the pixels equal to 1. Figure~\ref{psf_demo} shows that the hybrid PSF models show significant improvement outside $\sim 1\arcsec$, especially in long-wavelength filters redder than F200W.

\subsection{Simultaneous Multiwavelength Image and SED Fitting} \label{gs_fitting}

\subsubsection{Introduction to GalfitS}

From the standpoint of panchromatic image analysis, measuring AGN host galaxy properties involves two general steps. First, emission from the host galaxy needs to be separated from that of its central nucleus. This needs to be done for multi-band images, so as to construct host galaxy SEDs. Then, physical properties are derived from the observed SEDs, usually by fitting stellar population synthesis models combined with corresponding nebular emission templates. 

In terms of AGN-host decomposition, various tools are currently up for the task. Two-dimensional (2D) analysis codes such as {\tt\string Galfit} \citep{Peng2002,Peng2010} fit a set of parametric models to an image in a certain filter. In this case, the panchromatic SED for the host component is obtained by analyzing multiple bands separately. However, the morphology and structure of a galaxy do not vary independently across multiple bands. The wavelength dependence of galaxy morphology is determined by the spatial distribution of stellar population and dust, which means that the 2D model for the host galaxy in different bands should be interdependent. Galaxy structural parameters, such as the S\'ersic index and effective radius, vary smoothly and systematically with wavelength \citep[e.g.,][]{2012MNRAS.421.1007K, 2013MNRAS.435..623V, Treu2023, 2024ApJ...960..104S}. The host galaxy parameters across different bands can be linked by optimizing the parametric models in all bands simultaneously. Codes that have this capability include {\tt\string GalfitM} \citep{2013MNRAS.430..330H}, the multi-band extension of {\tt\string Galfit}, and {\tt\string GaLight} \citep{Ding2020}. Different priors are employed to constrain the wavelength variation of model parameters: {\tt\string GaLight} keeps the parameters constant across different bands, while {\tt\string GalfitM} allows their values to vary as a Chebyshev polynomial whose order can be set by the user. Despite the improvements compare to single-band analysis, there is still room for refinement, especially for the priors adopted. The wavelength variation of model parameters, particularly the integrated flux, are not arbitrary and should be consistent with physical expectations from stellar and nebular SED models. Instead of treating SED modeling as a subsequent and independent process after AGN-host decomposition (e.g., \citealt{Yu2024}), it should be incorporated into multi-band imaging analysis to provide physically motivated constraints. The effective radius or S\'ersic index of a galaxy generally vary smoothly and monotonically with wavelength, in accordance with observed trends in color gradients that reflect genuine, systematic variations in radius of the age, metallicy, or dust reddening (e.g., \citealt{Searle1973, Balcells1994, deJong1996}). Linking structural parameters across different bands using arbitrary or mathematical constraints can lead to unphysical results.

This study uses a {\tt\string GalfitS} (R. Li \& L. C. Ho, in preparation), a new open-source Python package designed for simultaneous multi-band image analysis. After supplying the multi-band images and error maps as input data, as well as their corresponding PSFs, the user defines a parametric profile for each component of the model, such as a a S\'ersic function for the galaxy and a point-source for the nucleus. Distinct from the other above-mentioned codes, {\tt\string GalfitS} assigns to each component a physically motivated SED model. For example, the galaxy component considers both stellar and nebular emissions, while active nucleus includes contributions from an accretion disk, dusty torus, and so forth. Subsequently, a model datacube is generated that encapsulates the spatial distribution of light and the SED derived from the theoretical models. Model images are generated by integrating the datacube along the wavelength dimension with the filter response curve and convolving with the PSF in each band. {\tt\string GalfitS} then calculates the likelihood function using $\chi^2$ statistics and performs Bayesian-based parameter optimization. To extract more meaningful morphological or SED parameters, the fit can be constrained by imposing additional astrophysical priors, such as the stellar mass-metallicity relation or the stellar mass-size relation (MSR). Once the user specifies the median relation and scatter of the astrophysical prior, an additional likelihood calculated from the deviation of the current parameter values relative to the input prior, normalized by its scatter, will be incorporated into the total likelihood function during the optimization process. 

{\tt\string GalfitS} uses a grid based on WCS coordinates instead of a pixel grid, such that, so long as accurate PSF models are provided, images from different instruments, or from the same instrument but with different pixel scales, can be fit simultaneously without rebinning them to a common pixel scale and convolving them to the same resolution. 
In this case, bands with higher spatial resolution can constrain the morphology, which may otherwise be unresolved or blended, while bands with lower resolution can still provide information on the total flux.
Several optimization algorithms are available, ranging from the classical \cite{Levenberg 1944}--\cite{Marquardt 1963} algorithm to dynamical nested sampling \citep{Higson2019}. We refer readers to R. Li \& L. C. Ho (in preparation) for details.

\begin{figure*}[ht!]
\centering
\includegraphics[width=0.75\textwidth]{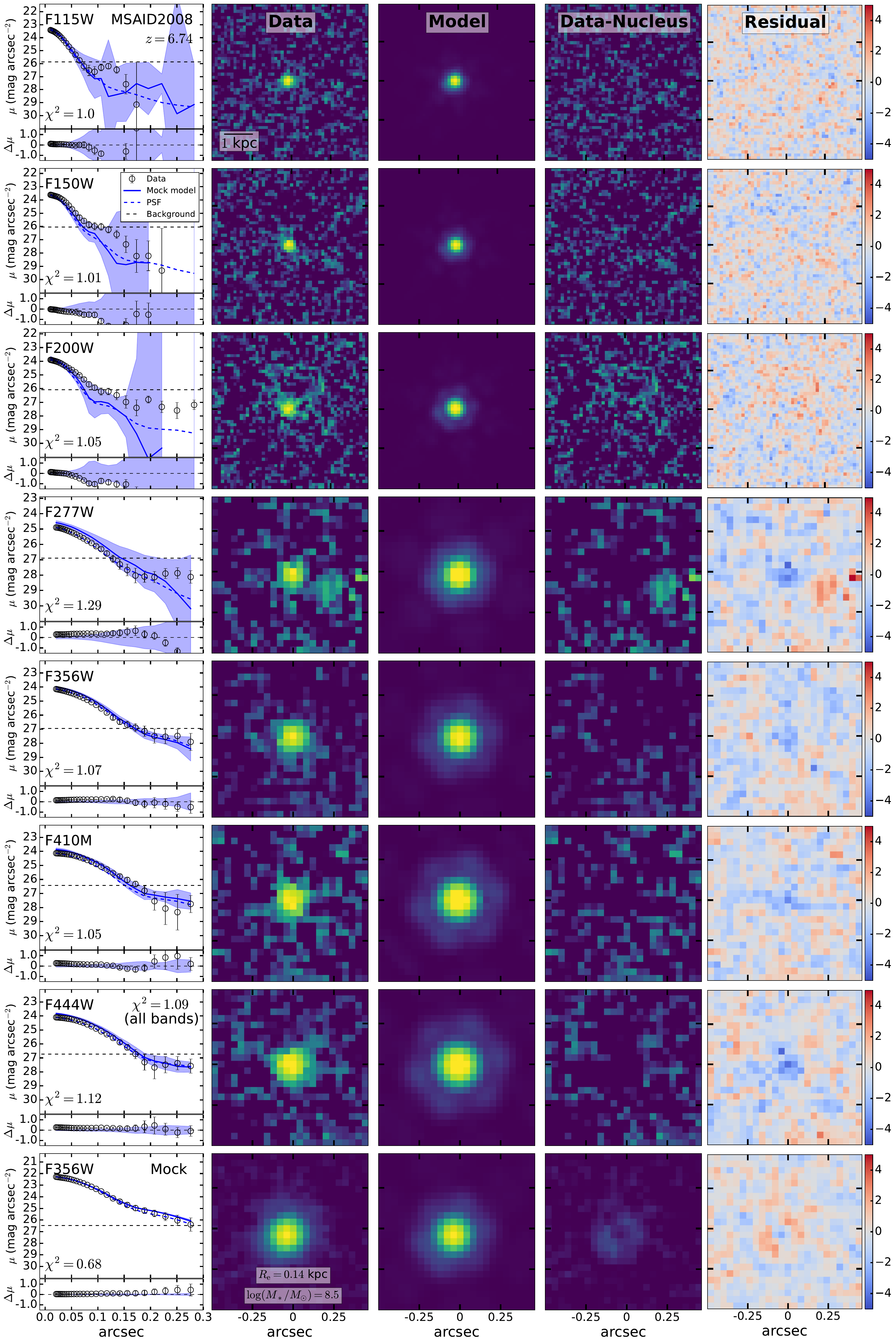}
\caption{Simultaneous multi-band image fitting results using a single point-source model for MSAID2008 at $z=6.74$. The first seven rows, from top to bottom, show the results for the seven NIRCam bands (F115W, F150W, F200W, F277W, F356W, F410M, F444W). In the left-most column, the upper panel of each row shows the radial surface brightness distribution (open circles with error bars), the PSF model (blue dotted line), as well as the median profile of the mock point source (blue solid line) and its standard deviation (blue shaded region). The background noise level is denoted by the black horizontal dashed line. The $\chi^2$ for each band is given in the lower-left corner of each panel, while that for all seven bands is given in the upper-right corner the panel for F444W. The lower subpanels give the residuals between the data and the best-fit model (data$-$model). The imaging columns, from left to right, display the original data, best-fit model, data minus the nucleus component, and residuals normalized by the errors (data$-$model$/$error), which are stretched linearly from $-5$ to 5. The bottom row illustrates the results of one of the mock images created for the target LRD, selected to reveal the host galaxy detection limit with the lowest stellar mass and smallest effective radius. Section~\ref{upper_limit} describes the mock simulations and detection limit calculation.
\label{2008_fitting_demo}}
\end{figure*}

\begin{figure*}[ht!]
\centering
\includegraphics[width=0.75\textwidth]{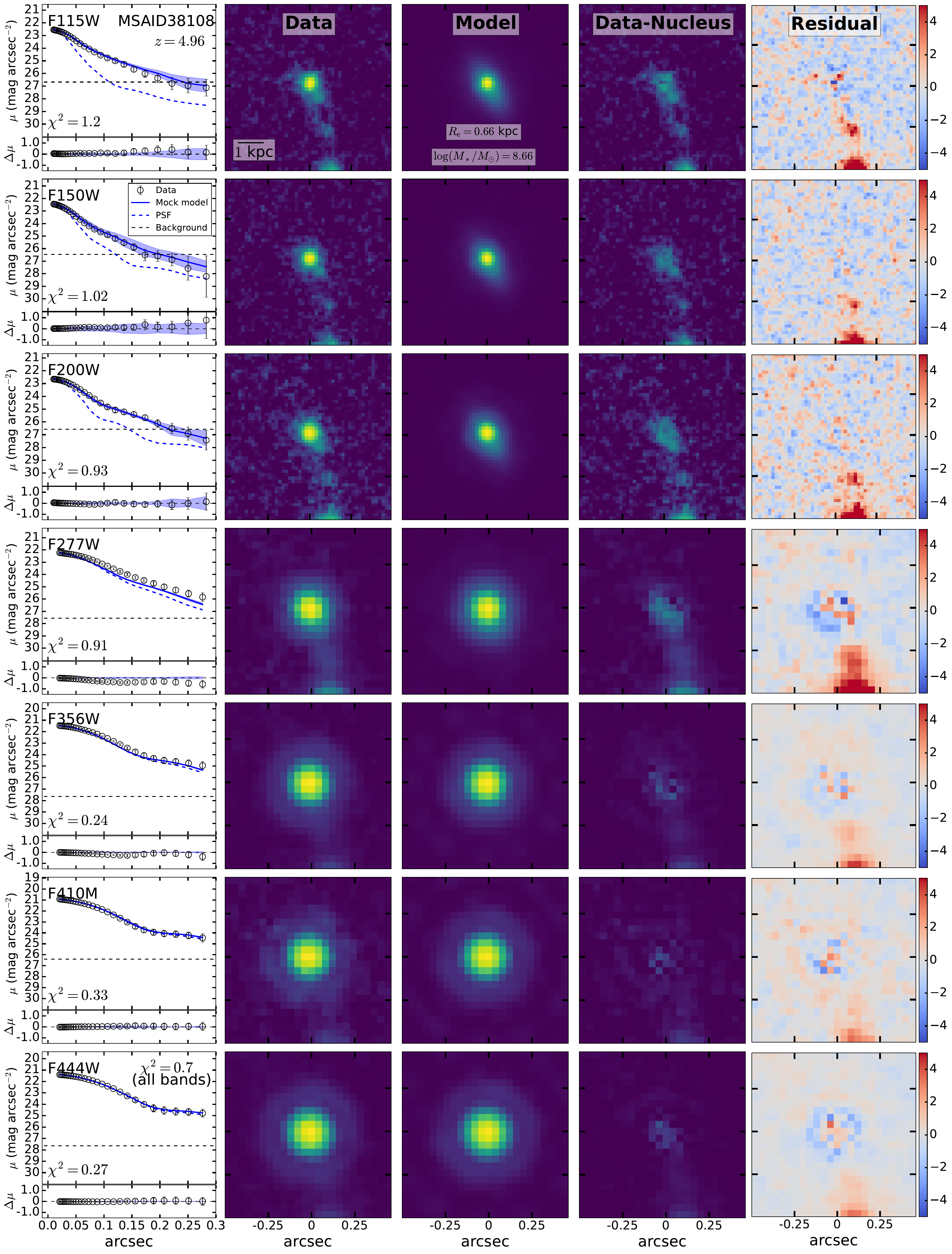}
\caption{As with Figure~\ref{2008_fitting_demo}, but for MSAID38108, which has detected extended emission centered on the point source.
\label{host_detected_demo}}
\end{figure*}

\begin{figure*}[ht]
\centering
\includegraphics[width=0.75\textwidth]{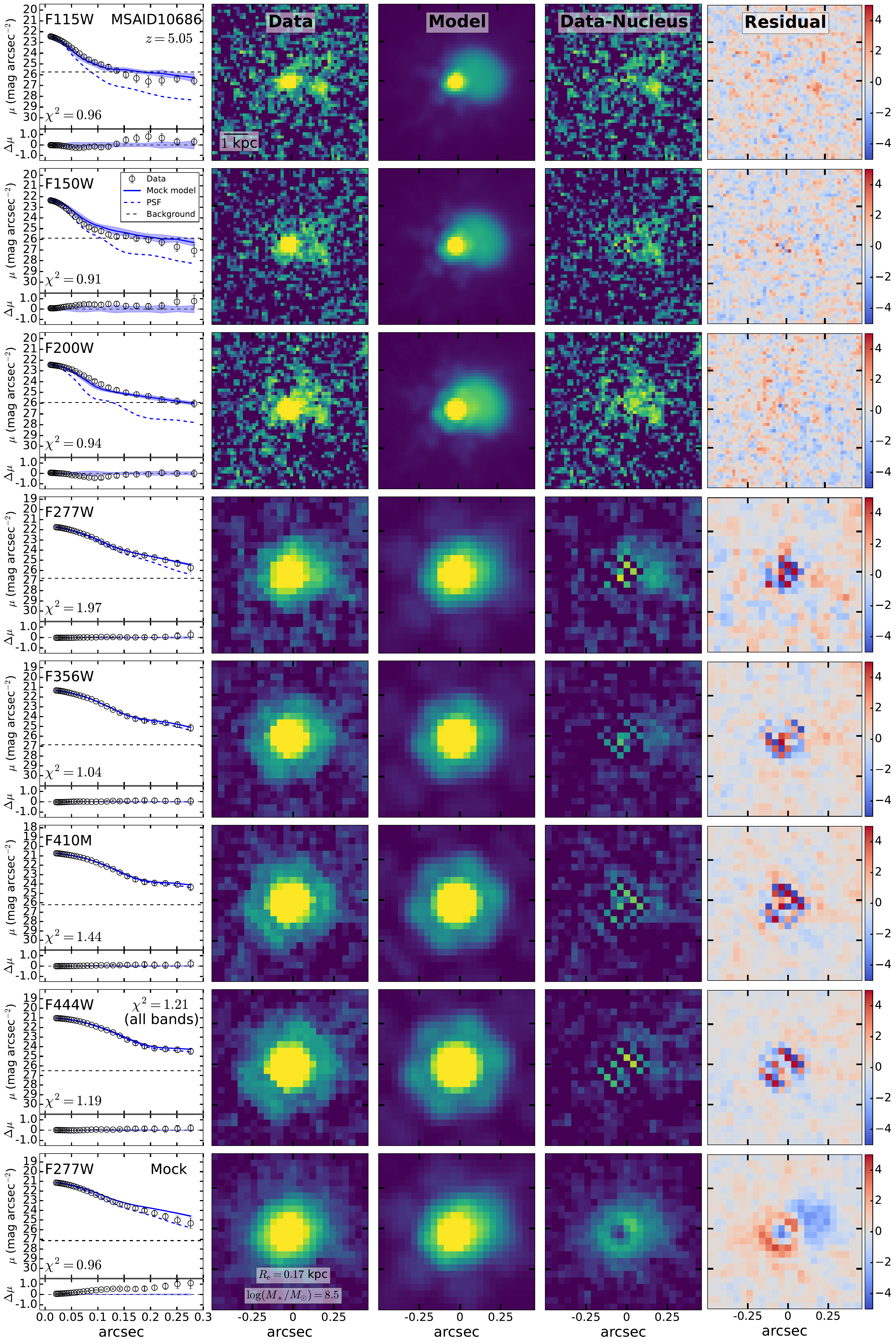}
\caption{As with Figure~\ref{2008_fitting_demo}, but for MSAID10686, which has off-centered emission.
\label{off-center_demo}}
\end{figure*}

\subsubsection{Data Preparation and Fitting Procedure}

For every LRD in our sample, we first create a $4\arcsec \times 4\arcsec$ cutout image in the F277W band, which serves as the detection image.
We adopt the F277W band for source detection, as the images exhibit less clutter while maintaining sufficient resolution to identify nearby sources, which are typically more prominent in the SW filters. The relatively large image size is chosen to ensure that we can estimate an accurate background fluctuation level, a key parameter in source detection. Source detection is then performed using the {\tt\string Photutils.detect\_source} package, with {\tt\string nstd}$=0.7$, {\tt\string npix=15}, and {\tt\string contrast}$=0.001$. We generate a mask image from the derived segmentation map by masking all nearby sources. To ensure that the area masked is consistent across different filters, for all bands we mask the same sky region as those masked in the detection image according to their WCS coordinates. With all the sources masked, the local sky background is then calculated and subtracted through sigma-clipping. Mosaics from UNCOVER DR2 do not contain error maps that account for the pixel-wise noise. For each band, we estimate two contributions to the error map: (1) Poisson noise, which is calculated from the science image and the effective exposure time assuming pure Poisson process; and (2) a constant error across the target cutouts representing an underlying Gaussian background, the value of which is taken as the sigma-clipped standard deviation. Each error map is then scaled so that after masking all sources the median value of the error map equals to the standard deviation of the mosaic. For the fitting procedure, the image cutout size is chosen to include all emission from the target while minimizing contamination from nearby sources. The final cutout image, error map, and mask image have a size of $1\arcsec\times 1\arcsec$ centered on the target LRD, for all sources except MSAID45924, for which we use a larger cutout size of $2\arcsec\times 2\arcsec$ to accomodate the prominent diffraction spikes in bands redder than F277W. 

We fit the host galaxy component using a \cite{Sersic1968} function,
 
\begin{equation}
\Sigma(R)=\Sigma_e\exp{\left\{-\kappa\left[\left(\frac{R}{R_e}\right)^{1/n}-1\right]\right\}},
\end{equation}

\noindent
where $R_e$ is the effective radius of the galaxy that contains half of the total flux, $\Sigma_e$ is the surface brightness at $R_e$, the index $n$ specifies the shape of the light profile, and $\kappa$ is related to $n$ by the incomplete gamma function, $\Gamma\left(2n\right)=2\gamma\left(2n,\kappa\right)$ \citep{1991A&A...249...99C}. The host galaxy component is associated with an SED model consisting of a stellar population, nebular emission, and dust attenuation. We adopt a non-parametric star formation history divided into four equal logarithmic age bins, whose width is determined by averaging the age of the Universe at the corresponding redshift of the source assuming star formation starts 200 Myr after the Big Bang. We employ {\tt\string STARBURST99} \citep{1999ApJS..123....3L} to generate simple stellar populations, assuming a \cite{Kroupa2001} stellar initial mass function and a sub-solar metallicity of $Z=0.004$, motivated by the typical chemical abundances of galaxies at $z\gtrsim 3$ \citep[e.g.,][]{Faisst2016, Jones2020, Langeroodi2023}. Nebular emission is interpolated from a library of theoretical spectra calculated using the photoionization code {\tt\string CLOUDY} (version 17.03; \citealt{Ferland2017}) assuming a range of metallicities and ionization parameters. 
We adopt the \citet{2000ApJ...533..682C} dust extinction curve, allowing the $V$-band attenuation ($A_V$) to be a free parameter\footnote{We extend the extinction curve between the Lyman break and 1500~\AA\ following \citet{2002ApJS..140..303L}, similar to the implementation in {\tt\string CIGALE} \citep{2019A&A...622A.103B}.}. 

The light profile of the AGN component is described by a point source to mimic the unresolved emission from the central nucleus. Instead of using empirical or analytical AGN SED templates \citep[e.g.,][]{2001AJ....122..549V, 2006ApJS..166..470R,2016MNRAS.458.2288S, 2018ApJ...866...92L, 2021MNRAS.508..737T} to describe the nuclear component, we allow the nuclear flux to be optimized independently across different bands, without any constraints on the spectral shape. Conventional AGN templates, although capable of accounting for the observed SEDs of LRDs in the rest-frame UV/optical using a combination of obscured and scattered AGN emission (e.g., \citealt{2024ApJ...964...39G}), predict strong emission from the hot, dusty torus, in apparent conflict with the relatively weak rest-frame near-infrared emission actually observed \citep{2024ApJ...968....4P, 2024arXiv240302304W, 2024ApJ...968...34W}. But perhaps conventional AGN templates should not be applied to LRDs. After all, no known AGN classes below $z \approx 4$ can match the LRDs in terms of their SED, host morphology, and number density (Section~1). The physical properties of the dust \citep{2024arXiv240710760L} and gas \citep{2024arXiv240907805I} also appear to be unique in these systems. In light of these considerations, we choose not not to impose any priors on the AGN component, apart from requiring that the source be spatially unresolved, hoping to provide an independent measurement of the nuclear SED.

We fit two sets of models for each LRD: one consisting of a nucleus plus a host galaxy, represented by a point source at the center of a S\'ersic profile, and the other only the nucleus. A central point source is included in both sets of models because the presence of broad H$\alpha$ emission strongly suggests that there is a relatively unobscured AGN (but see \citealt{2024arXiv240807745B}). The overall consistency between the observed luminosity of the 5100\,\AA\ continuum (de-reddened by fitting an obscured AGN model to the rest-frame optical SED) and the same value inferred from broad H$\alpha$ emission using the empirical relation established for low-redshift AGNs \citep{2005ApJ...630..122G} supports the notion that a significant fraction, if not all, of the rest-frame optical continuum arises from nonstellar emission from BH accretion \citep{2024ApJ...964...39G}.  Indeed, equivalent widths of broad H$\alpha$ emission among high-redshift AGNs discovered by JWST are, if anything, {\it higher}\ (by a factor of $\sim 2-3$) than observed in low-redshift AGNs \citep{2024arXiv240500504M}, consistent with theoretical predictions of rapidly growing seed BHs at high redshift \citep{Inayoshi2022} and precisely the opposite of what is expected if stars contribute substantially to the optical continuum.

Whenever a host galaxy component is included in the fit, we link the position of the point source with the center of the S\'ersic profile. We do not allow for any displacement between the nucleus from its host galaxy, for neither gravitational recoil \citep{1973ApJ...183..657B, 2007ApJ...659L...5C} nor gravitational slingshot involving triple supermassive BHs \citep{2007MNRAS.377..957H} is likely. These processes generate recoil velocities of hundreds up to a thousand $\rm km~s^{-1}$, which would be imprinted as significantly blueshifted or asymmetric broad emission lines, neither of which has yet been seen in our sample. Another possibility for displacement of the nucleus is an ongoing merger, which would shift the flux-weighted center away from the supermassive BH. Indeed, we observe extended emission near the point source in 5 out of the 8 LRDs in our sample. Their effect on the host galaxy mass estimate will be discussed in Section~\ref{discussion}.

We evaluate the goodness-of-fit by the Bayesian information criterion

\begin{equation}
{\rm BIC} =k\, {\rm ln}(n)-2\,{\rm ln}(L),
\end{equation}

\noindent
in which $k$ is the number of model parameters, $n$ is the number of data points, and $L$ is the fitting likelihood. A smaller BIC value indicates better fitting performance.

\begin{deluxetable*}{cccccccccccc}
\label{extended_table}
\tablecaption{Photometry and Morphological Parameters of LRDs with Extended Emission}
\tablewidth{0pt}
\tablehead{MSAID & F115W & F150W & F200W & F277W & F356W & F410M & F444W & $z_{\rm comp}$ & $R_e$ & $n$ & $d_{\rm cen}$ \\
                    & (mag) & (mag)& (mag)& (mag)& (mag)& (mag)& (mag) & & (kpc) & & (kpc) \\
               (1) & (2) & (3) & (4) & (5) & (6) & (7) & (8) & (9) & (10) & (11) & (12)}
\startdata    \label{table:results}
4286  & 28.61$^{+0.15}_{-0.12}$ & 28.85$^{+0.09}_{-0.08}$ & 28.82$^{+0.09}_{-0.08}$ & 28.47$^{+0.11}_{-0.09}$ & 28.10$^{+0.11}_{-0.09}$ & \nodata                 & 28.02$^{+0.14}_{-0.14}$ & 5.84 & $0.29^{+0.04}_{-0.03}$ & $0.20^{+0.36}_{-0.33}$ & $0.74^{+0.12}_{-0.08}$ \\
10686 & 28.16$^{+0.07}_{-0.07}$ & 28.01$^{+0.06}_{-0.06}$ & 27.77$^{+0.05}_{-0.05}$ & 27.54$^{+0.07}_{-0.07}$ & 28.37$^{+0.05}_{-0.06}$ & $>28.05$                & $>28.23$                & 5.05 & $0.79^{+0.21}_{-0.21}$ & $0.20^{+0.01}_{-0.01}$ & $1.21^{+0.05}_{-0.05}$ \\
13821 & 28.64$^{+0.08}_{-0.06}$ & 28.48$^{+0.07}_{-0.06}$ & 28.44$^{+0.07}_{-0.06}$ & 28.75$^{+0.07}_{-0.07}$ & 27.77$^{+0.05}_{-0.05}$ & $>29.51$                & 27.77$^{+0.06}_{-0.06}$ & 6.34 & $0.36^{+0.12}_{-0.07}$ & $1.70^{+0.05}_{-0.08}$ & $0.38^{+0.01}_{-0.01}$ \\
38108\tablenotemark{a} & 27.84$^{+0.17}_{-0.13} $                & 27.94$^{+0.16}_{-0.11}$                   & 27.94$^{+0.16}_{-0.09}$                      & 27.42$^{+0.15}_{-0.07}$                     & 27.57$^{+0.13}_{-0.06}$                      &  27.53$^{+0.12}_{-0.06}$                     & 27.65$^{+0.12}_{-0.05}$                      & 4.96 & $0.66^{+0.08}_{-0.05}$ & $0.71^{+0.07}_{-0.08}$ & fixed  \\
41225 & 28.22$^{+0.06}_{-0.06}$ & 28.07$^{+0.05}_{-0.05}$ & 27.60$^{+0.04}_{-0.04}$ & 27.72$^{+0.05}_{-0.05}$ & 27.78$^{+0.05}_{-0.05}$ & 27.73$^{+0.05}_{-0.05}$ & 27.70$^{+0.06}_{-0.06}$ & 2.69 & $0.71^{+0.04}_{-0.03}$ & $0.98^{+0.06}_{-0.07}$ & \nodata   \\
45924 & 27.24$^{+0.10}_{-0.06}$ & 27.55$^{+0.06}_{-0.04}$ & 27.08$^{+0.06}_{-0.04}$ & $>33.10$                & $>30.76$                & $>32.49$                & $>30.82$                & 4.46 & $1.14^{+0.10}_{-0.05}$ & $0.73^{+0.20}_{-0.10}$ & $0.94^{+0.03}_{-0.04}$ \\
\enddata
\tablecomments{Col. (1): Object ID in MSA configuration \citep{2024ApJ...964...39G}. Cols. (2)$-$(8): Integrated magnitude and associated uncertainties estimated from the posterior distribution.
Col. (9): Redshift for the extended component; for MSAID41225, the redshift is taken as $z_{\rm ml}$ calculated from {\tt\string EAzY} (Section~\ref{off-center_discussion}), while for the other targets the redshift is the same as the spectroscopic redshift in Table~\ref{results_table}. Col. (10): Effective radius. Col. (11): S\'ersic index. Col. (12): Physical distance of the center of the extended component to the central point source.
\tablenotetext{a}{The extended emission in MSAID38108 is concentric with the point source and is considered to be stellar emission from the host galaxy. The extended component of the other sources is off-centered; their nature is discussed in Section~\ref{off-center_discussion}}.
}
\end{deluxetable*}

The spatial resolution of the UNCOVER DR2 mosaics is higher at shorter wavelengths, both in terms of the sharpness of the PSF \citep[$\rm FWHM = 0\farcs06$ in F115W and $0\farcs16$ in F444W;][]{2024ApJ...962..139Z} and pixel scale ($0\farcs02$ in SW filters and $0\farcs04$ in LW filters). As a result, the SW filters contain most of the morphological information, while the LW filters provide constraints on the total flux. Therefore, when fitting the data with the AGN$+$host model, we proceed in two steps. We initially only fit the three SW bands (F115W, F150W, F200W) with the aim of obtaining a rough, initial depiction of the host galaxy light profile. Then we subsequently fit all bands simultaneously, using the results from the first step as initial value so as to model the full SED of the host galaxy and its physical properties. The initial value of the effective radius $R_e$, position angle, and axis ratio of the S\'ersic profile are estimated from the segmentation map. Initial value of host galaxy stellar mass is calculated from the segmentation flux in rest-frame $B$ band assuming a mass-to-light ratio of 0.1, which roughly equals the observed value for high-redshift galaxies in the UNCOVER survey (Section~\ref{upper_limit}). We also notice that the final result is highly sensitive to the initial value of dust attenuation and S\'ersic index. In order to fully map the parameter space, we run multiple fittings in the first step with different combinations of initial values for $A_V \approx 0-2$ mag and $n\approx 1-4$. We use the intial result with the smallest BIC value as input for step two. 

We also explore the effect of constraining the host galaxy with the mass-size relation, using the parameterization for high-redshift galaxies \citep{2014ApJ...788...28V}

\begin{equation}
R_e=A\left(M_*/5\times 10^{10}\,M_{\odot}\right)^{\alpha}\,\rm kpc.
\end{equation}

\noindent
\cite{2024ApJ...962..176W} measured the mass-size relation at $3.0\lesssim z \lesssim 5.5$ for star-forming galaxies with $M_*>10^{9.5}\,M_{\odot}$ in the CEERS field \citep{2023ApJ...946L..13F}. Using combined photomety and imaging data from HST and JWST from $\sim 0.6-4.5\,\mu {\rm m}$, they estimated stellar mass from SED fitting and measured galaxy effective radius at rest-frame 5000\,\AA\ by simultaneously fitting multi-band images with {\tt\string GalfitM}. We adopt their best-fit slope of $\alpha=0.25$ and their derived redshift evolution of effective radius at rest-frame 5000\,\AA, for a fixed stellar mass of $5\times 10^{10}\,M_{\odot}$,

\begin{equation}
\overline{R_e}=7.1\left(1+z\right)^{-0.63}\,\rm kpc.
\end{equation}

\noindent
The scatter is taken as a constant value of $\Delta \log\, (R_e/{\rm kpc}) = 0.3$. We use the mass-size relation at the corresponding redshift of each LRD.

\begin{figure*}[ht]
\centering
\includegraphics[width=\textwidth]{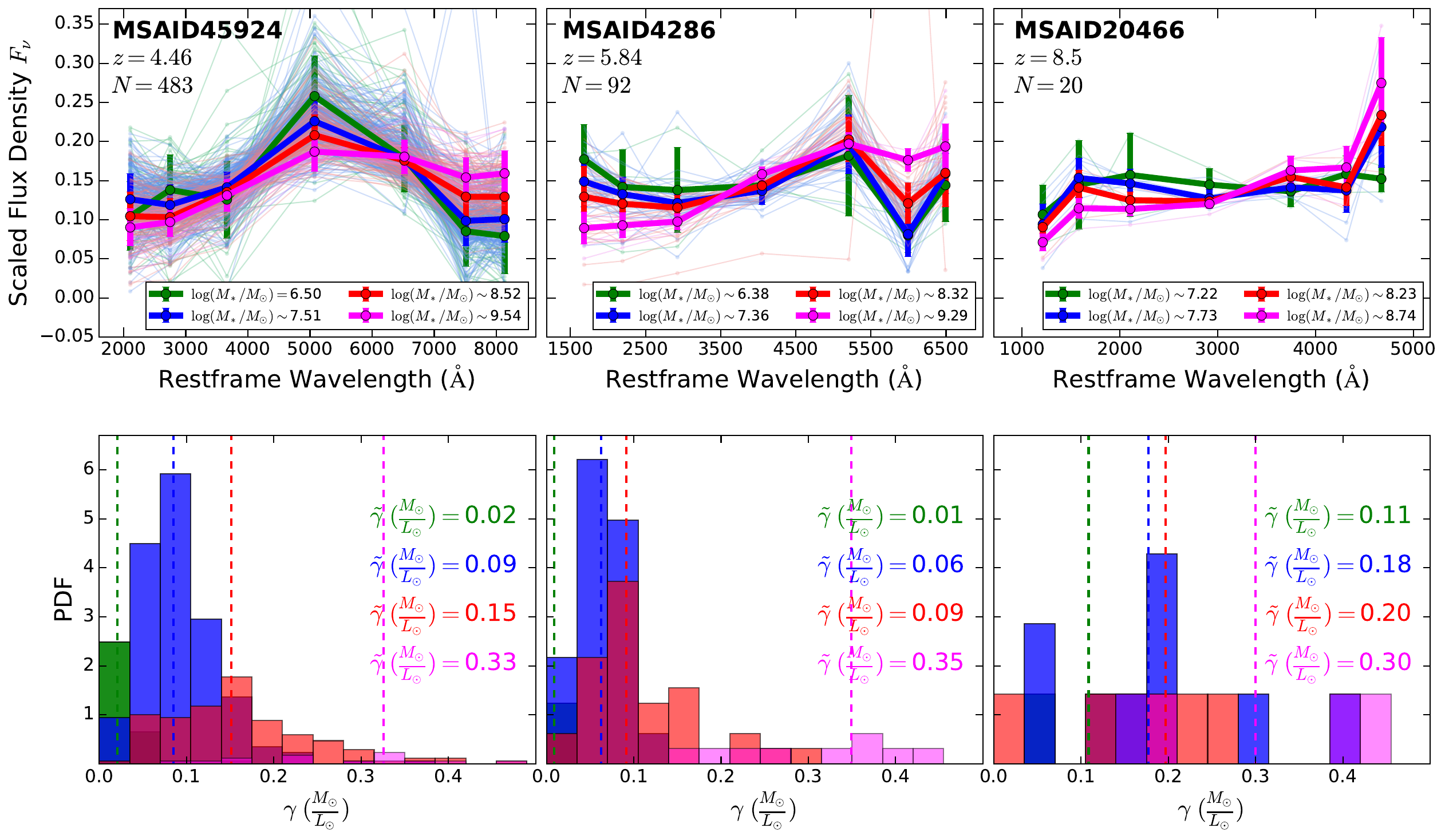}
\caption{The SED (top) and the distribution of rest-frame optical mass-to-light ratio (bottom) for the galaxies used in mock simulations for three LRDs in our sample. All results are plotted in four mass bins. In the top row, thinner lines are SEDs for individual galaxies normalized by their total flux in seven NIRCam bands, while their median SED is plotted as dots connected by thicker lines, and the error bars represent standard deviation. The total number of galaxies within $\Delta z<0.5$ selected for each LRD is labeled on the top-left corner of each panel. The bottom row plots the probability distribution function (PDF) of rest-frame optical mass-to-light ratio for the same group of galaxies, plotted with the same color as in the top ros. Verticle dashed lines gives the median value in each mass bin.
\label{sed_msr_demo}}
\end{figure*}

We observe extended, off-centered emission near the central point source in five LRDs (MSAID4286, 10686, 13821, 41225, and 45924). In order to measure the multi-band fluxes of these off-centered blobs, we refit the images with a point source to account for the nucleus in conjunction with a S\'ersic component to describe the off-centered emission. Since we do not know the physical nature of these blobs, we adopt a pure morphological fitting approach without adding SED models for constraints. Specifically, we assume that the underlying S\'ersic profile has constant effective radius, S\'ersic index, axis ratio, and position angle across all filters, allowing only the total flux to change in different bands. To determine whether the off-centered emission is robustly detected, we calculate the best-fit BIC values for single-band images in each filter before and after including the extra component. We treat the best-fit flux in each band as a reliable detection whenever the BIC value decreases after including the extra component; otherwise, the best-fit value is taken as a flux upper limit.

The fit uses the Adam Optimizer as the optimization algorithm to locate the position of maximum likelihood in parameter space. We run this process 15000 times. We then perform nested sampling around the position of maximum likelihood to infer parameter uncertainties from the posterior distribution.

\section{Results}\label{results}

Our fitting results using different models and priors are presented in Table~\ref{results_table}. For seven out of the eight LRDs in our sample, a single point-source always gives the best fit compared to including an additional host galaxy component. They show no detectable extended concentric emission that can be ascribed to the host galaxy. This general result is not affected by including the mass-size relation as a prior constraint in the fit. Figure~\ref{2008_fitting_demo} gives the 1D and 2D surface brightness distribution of the best-fit single point-source model for the sample source MSAID2008 at $z=6.74$. Fits for the other six sources are presented in Figures~\ref{off-center_demo} and A1--A5 (Appendix~\ref{app_a}). We do not merely use the best-fit model images when evaluating the 1D profiles. Instead, we create mock images by adding the model image in each band on top of empty sky regions taken from the mosaics, thereby creating a more robust realization of the uncertainties associated with the actual sky background fluctuations. For each LRD, the same procedure is repeated 10 times using different sky regions, and the median and standard deviation of the profiles for these mock model images in each band are taken as the best-fit model and its associated uncertainty. Our methodology is motivated by the fact that many of the LRDs are extremely faint in the SW filters ($\sim 29$ mag), nearly as faint as the background fluctuation level. Under these circumstances, correlated background fluctuations can be mistaken for real signal to produce slightly extended excess emission beyond an intrinsically point-like core. As can be seen in Figure~\ref{2008_fitting_demo}, the smooth residuals as well as the close resemblance to the radial profile of a pure point source (PSF) demonstrates the source is indeed completely unresolved. 

Extended emission in excess of a central point source is detected in six objects (Table~\ref{extended_table}). However, only one (MSAID38108 at $z = 4.96$) can be ascribed to host galaxy emission, if we require that the underlying extended component be centered on the point source (Figure~\ref{host_detected_demo}). As shown in Table~\ref{results_table}, constraining the S\'ersic profile with the mass-size relation yields the largest improvement in the statistical significance of the fit, with $\Delta \rm BIC=1838.36$, indicating a solid detection. The extended emission in the other objects is off-centered, as illustrated in Figure~\ref{off-center_demo} for the case of MSAID10686; other cases are shown in Appendix~\ref{app_a}. The possible nature of this emission is discussed in Section~\ref{off-center_discussion}.

\begin{figure*}[ht]
\centering
\includegraphics[width=\textwidth]{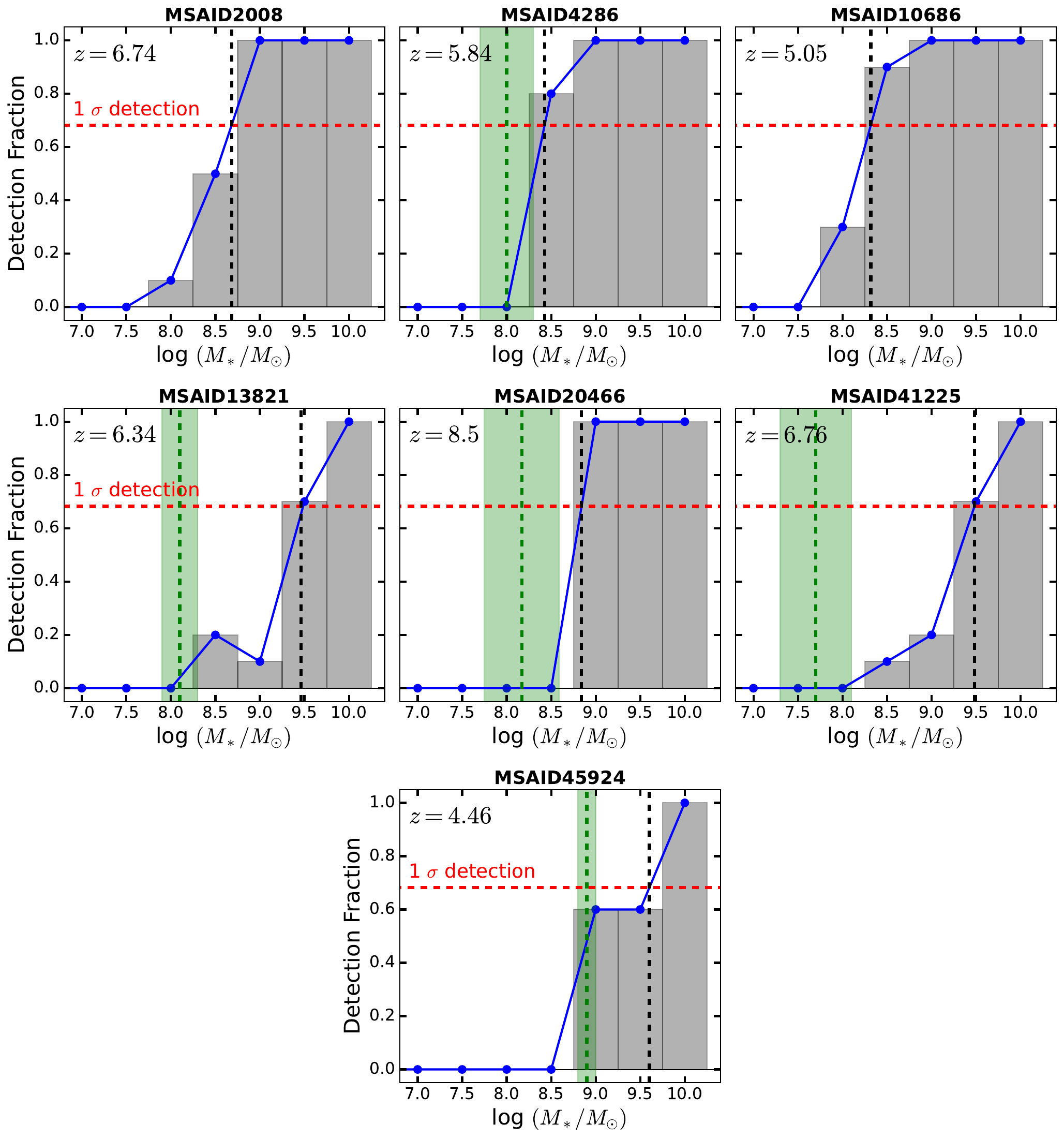}
\caption{Host galaxy detection probability as a function of stellar mass for the seven LRDs in our sample. In each panel, the red horizontal line represents the $1\, \sigma$ ($68.2\%$) detection rate. The detection probability in each stellar mass bin is shown by the grey histograms connected by the blue solid line. The vertical dashed line in each panel marks the stellar mass upper limit for each LRD, defined by the detection probability of $1\,\sigma$. Single-epoch virial BH mass is available for five of the sources \citep{2024ApJ...964...39G}, plotted as a vertical green dashed line, with light green shaded region indicating the measurement uncertainty.
\label{det_frac_demo}}
\end{figure*}

\begin{figure*}[ht!]
\centering
\includegraphics[width=\textwidth]{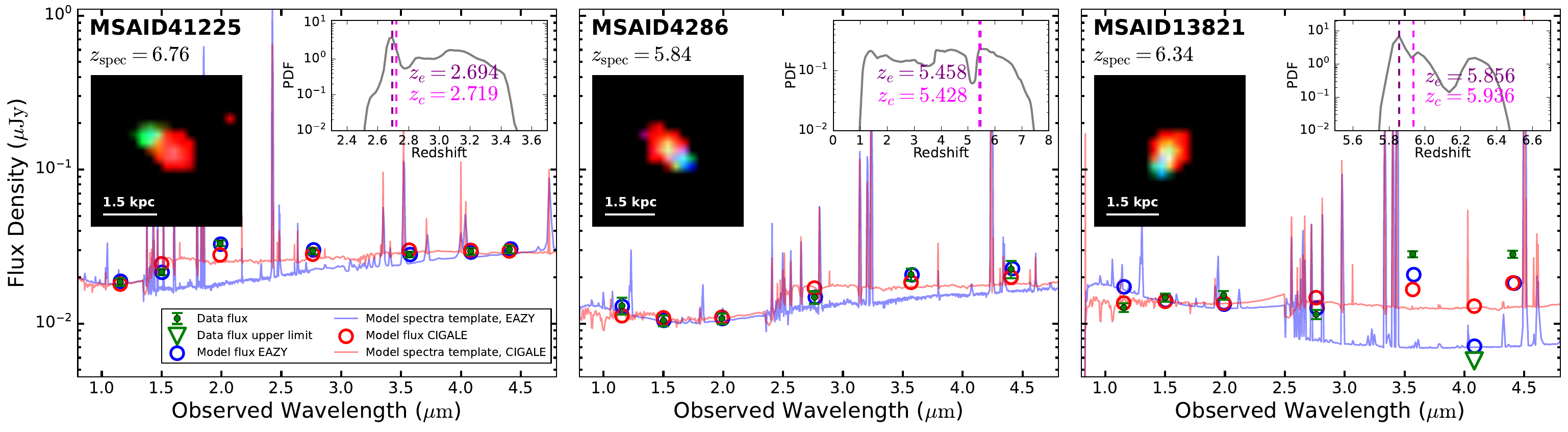}
\caption{Best-fit SED models from {\tt\string EAzY} (blue line) and {\tt\string CIGALE} (red line) for the off-centered emission in MSAID41225, 4286, and 13821. The measured fluxes for the solid detections are presented in filled green dots with error bars, while upper limits are marked as green triangles. The spectroscopic redshift for the LRD point source ($z_{\rm spec}$) is given in each panel. The inset panel on the top-right corner shows the probability distribution function (PDF) of redshift derived by {\tt\string EAzY}; the best-fit redshift from {\tt\string EAzY} ($z_e$) and {\tt\string CIGALE} ($z_c$) are denoted by the brown and magenta dashed line, respectively. The colored image for each source is generated based on cutouts in F115W (blue), F200W (green), and F444W (red); the fluxes from the three bands are rescaled in order to better highlight the off-centered blobs. 
\label{off-center_sed_phot-z}}
\end{figure*}

\subsection{Upper Limits on Stellar Mass of the Host Galaxy} \label{upper_limit}

As the host galaxy is undetected in seven out of eight LRDs, we attempt to set an upper limit on its stellar mass. We create realistic mock data by inserting images of high-redshift galaxies under the observed emission of the LRD point source, and then carefully analyze them to constrain the value of stellar mass above which the host galaxy would have been detected. We begin by selecting high-redshift galaxies in the UNCOVER field suitable for this purpose drawn from the catalog of \citet{2024ApJS..270...12W}, who provide physical properties, such as photometric redshift, stellar mass, and AGN fraction for all the photometrically identified galaxies \citep{2024ApJS..270....7W}. Spectroscopic redshift is also reported, when available. Our mock simulations are performed with a host sample derived from these high-redshift galaxies, assuming that the hypothetical host galaxies of the LRDs are drawn from the general galaxy population at the same redshift. 

We first select galaxies that have $z>4$, using spectroscopic redshifts when available, otherwise adopting the 50th percentile of the photometric redshift ({\tt\string z\_50}) and imposing the additional requirement that the difference between the 84th ({\tt\string z\_84}) and 16th ({\tt\string z\_16}) percentile be less than 1, so as to remove any objects that may have suffered catastrophic failure during the photometric redshift inference. To mitigate against possible AGN contamination, we also limit the 50th percentile of the inferred logarithmic AGN contribution ({\tt\string logfagn\_50}) to $< 0.1$. Since all the LRDs in our sample are selected to be far from the center of the Abell~2744 cluster, we use the lensing model of \citet{2023MNRAS.523.4568F} to apply the same lensing magnification constraint ($\mu <2$) to ensure that the galaxies come from the same sky area as the LRDs. A total of 2345 galaxies are selected for our mock simulations, spanning stellar mass $M_*\approx 10^6-10^{11}\,M_{\odot}$ at a median redshift of $z = 4.8$.

Our ability to detect the host galaxy component depends on various factors. These include the luminosity and structural parameters, which determine the overall brightness and extendedness of the galaxy. Moreover, because the spatial resolution differs among the different bands, host galaxy detection using simultaneous multi-band decomposition is also affected by the galaxy SED. In principle, all these factors need to be taken into account in the mock simulations to best mimic the galaxy population at high redshift. In practice, however, the interdependence between panchromatic morphology and stellar population further complicates the situation. Galaxy size correlates with stellar mass, and the correlation varies with wavelength according to the spatial distribution of stars and dust. The morphology of a galaxy appears clumpier and more irregular in the rest-frame UV than in the optical \citep[e.g.,][]{1995ApJ...451L...1S, 2007ApJ...659..162T}, an effect difficult to be described by a smooth parametic surface brightness distribution, especially high redshifts \citep[e.g.,][]{1995ApJ...449L..23D, 2004ApJ...600L.139C, 2018ApJ...853..108G}. Detailed exploration of the impact of wavelength variation in galaxy size and substructure are beyond the scope of this paper. For the current simulations, we only create mock images in a single band whose effective wavelength is closest to the rest-frame optical (5000\,\AA), in which the stellar morphology is typically relatively smooth (e.g., \citealt{Ho2011}). This wavelength is also chosen to match the measurement of the mass-size relation \citep{2024ApJ...962..176W} adopted in our fitting and mock simulation.

\begin{figure*}[ht!]
\centering
\includegraphics[width=\textwidth]{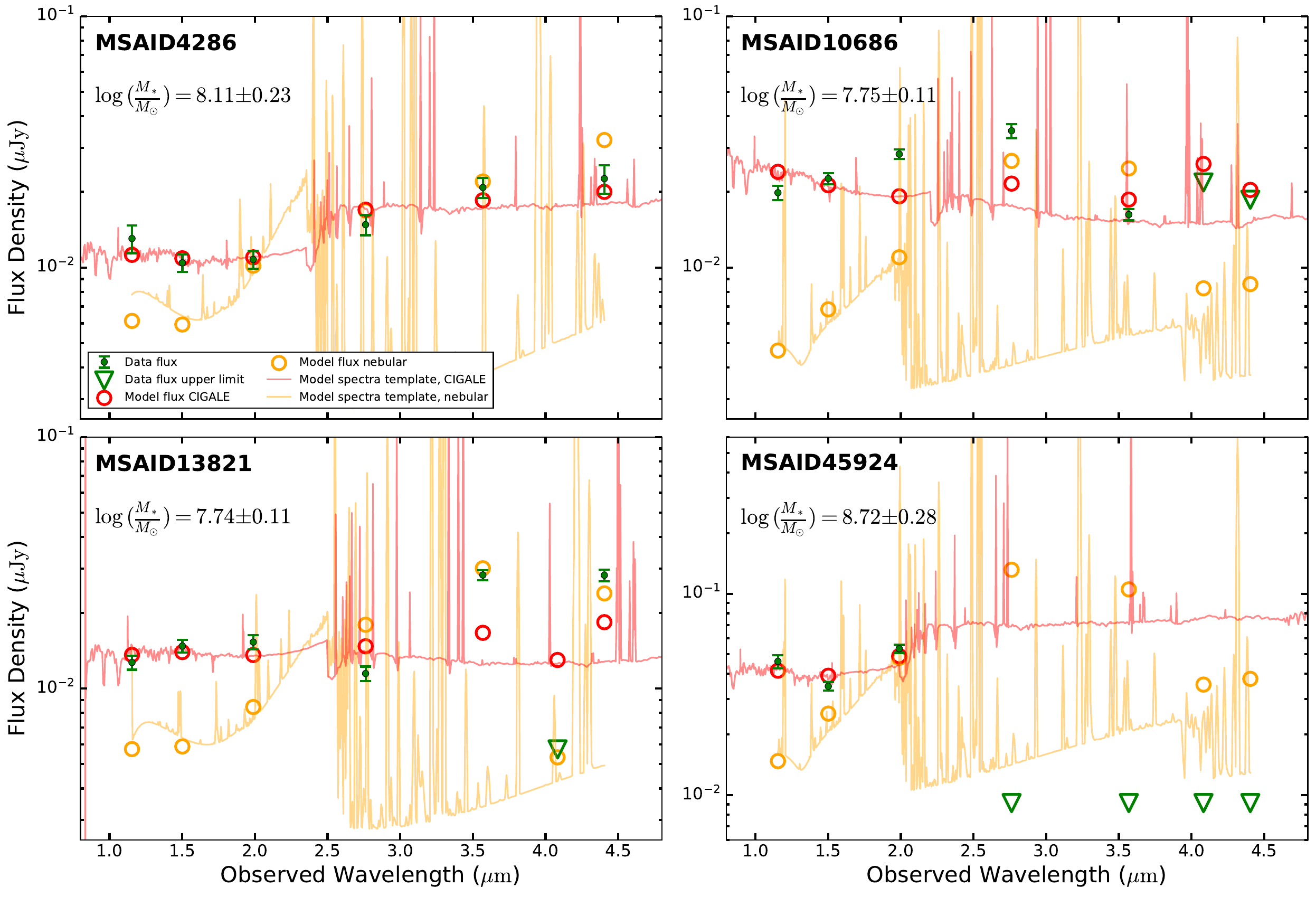}
\caption{Same as Figure~\ref{off-center_sed_phot-z}, but for best-fit SED models from {\tt\string CIGALE} (red line) and pure nebular emission templates (orange line) for the off-centered emission in MSAID4286, 10686, 13821 and 45924. The best-fit stellar mass is given in each panel. Note that for MSAID45924, the flux upper limits of the four bands (F277W, F356W, F410M, and F444W) are plotted to be the same ($m_{\rm AB}=29$ mag) for the purposes of the display; the actual values are reported in Table~\ref{extended_table}.
\label{off-center_sed_measurement}}
\end{figure*}

For each source in our sample, we select galaxies whose redshift differs by less than 0.5 from the LRD redshift. We use a S\'ersic function to represent the host galaxy profile. For a given stellar mass, the effective radius $R_e$ is drawn randomly from a log-normal distribution whose mean and standard deviation are calculated from the mass-size relation at the LRD redshift, as described in Section~\ref{gs_fitting}. The S\'ersic index $n$ and axis ratio are selected from the distribution of high-redshift ($4 < z < 9$) galaxies measured in \cite{2024ApJ...960..104S}, and the position angle can vary from 0 to 180 degrees. To set the flux for a given stellar mass, we randomly choose a galaxy in our sample and calculate the expected brightness using its mass-to-light ratio. 

The mass-to-light ratio is strongly affected by the specific star formation rate of the galaxy, which depends on its stellar mass \citep[e.g.,][]{2010ApJ...721..193P}. Therefore, for each LRD in our sample, we evenly split the galaxies at similar redshift into four logarithmic mass bins, and the galaxies from which we derive the mass-to-light ratios are chosen in the corresponding mass bin of the given stellar mass. Figure~\ref{sed_msr_demo} shows the SEDs and the mass-to-light ratio distributions of the galaxies used to create the mock images. For each of the three LRDs illustrated in the figure, the selected galaxies at the same redshift show similar spectra, with massive galaxies showing slightly redder colors and higher mass-to-light ratios than less massive ones, reflecting the more evolved stellar population in massive galaxies. The low overall optical mass-to-light ratios in these galaxies---$\gamma \lesssim 0.4\, M_{\odot}/L_{\odot}$---resembles that of local dwarf irregular galaxies \citep{1979ARA&A..17..135F, Herrmann2016}, which are experiencing ongoing star formation. The final single-band mock images are generated by combining the image of the PSF-convolved S\'ersic profile, a point source with the same flux as the LRD, and a visually selected empty sky region from the UNCOVER mosaics that has the same on-sky position angle combination as the target, and random pixel-to-pixel Poisson noise calculated from the total flux of all the afore-mentioned components. We require that the surface brightness fluctuation level of the empty sky region differs by less than $0.5\, {\rm mag\,arcsec}^{-2}$ compared with the background near the target, to ensure that the mock image is truly an analog to the observed data.
 
The mock LRD images are initially created with host stellar mass $M_* = 10^6 \, M_{\odot}$, which is gradually raised in steps of $\Delta \log\, (M_*/M_{\odot})=0.5$. For each stellar mass, 10 mock LRDs are produced with different structural parameters and mass-to-light ratios. Since we only conduct mock simulations for the rest-frame optical band, we use the pure morphological fitting setup described in Section~\ref{gs_fitting}, with no SED information included when fitting the single-band mock images. We compare the pure point-source fitting and joint point-source$+$S\'ersic fitting, and mark the host galaxy as detected once the BIC of the latter becomes smaller than that of the former. To simulate the effects of PSF mismatch, we create two PSF models, each using about half of the stars observed with the same telescope position angle as the target LRD. One of the PSFs is used to generate the mock LRD image, while the other is used for the fitting.

Figure~\ref{det_frac_demo} presents the host galaxy detection probability as a function of host galaxy stellar mass, defined as the number of mock LRDs with their host galaxy detected relative to the total number of mocks created in a single stellar mass bin. The trend increases with stellar mass, as expected. We define the $1\,\sigma$ upper limit for the host galaxy stellar mass as the value where detection probability reaches $68.2\%$. The exact number is calculated by linearly interpolating between the two adjacent stellar mass bins whose detection probabilities are below and above $1\,\sigma$. Overplotted in green are the single-epoch BH mass estimates \cite{2024ApJ...964...39G} for five of the LRDs (MSAID4286, 13821, 20466, 38108, and 45924). In the last row of Figures~\ref{2008_fitting_demo} and \ref{off-center_demo}, we present the fitting results for one of the mock images created for MSAID2008 and MSAID10686. These mock images were selected to represent the detection limits for host galaxies with the lowest stellar mass and smallest effective radius in our sample. Fitting results for the remaining five sources are provided in Figures A1--A5. When compared to real images in the same band, the mock images are generally brighter due to the inclusion of additional mock galaxies beneath the primary sources. These mock images are used to estimate the upper limit of the host galaxy mass based solely on unresolved morphology. The mass upper limit derived from the total observed flux is discussed in Section~\ref{discussion}.

\subsection{Nature of the Off-centered Emission} \label{off-center_discussion}

We perform SED fitting to explore the origin of the off-centered emission detected in five of the LRDs (Table~\ref{extended_table}). We attempt to estimate the photometric redshift ($z_{\rm phot}$) of the three sources (MSAID4286, 13821, and 41225) whose off-centered flux was detected in at least six bands. We compute $z_{\rm phot}$ with the code {\tt\string EAzY} \citep{2008ApJ...686.1503B} using 12 stellar population synthesis SED templates. To allow for the possibility that the off-centered emission originates at redshift similar to or even higher than the LRD, we also include templates from \citet{2023ApJ...958..141L} that have been optimized for galaxies at $z>8$. Magnitude prior is turned off during the fitting, and the redshift range is set from 0 to 12, with step size 0.005. To verify the results from {\tt\string EAzY} and to derive other physical properties for sources, we use {\tt\string CIGALE} \citep[]{2019A&A...622A.103B, 2020MNRAS.491..740Y} to fit the detected flux as well as upper limits in all seven bands, using the redshift with the highest posterior probability from {\tt\string EAzY}, $z_{\rm ml}$, as initial redshift value. We adopt a commonly used delayed-$\tau$ module ({\tt\string sfhdelayed}) for the star formation history and employ \citet{2003MNRAS.344.1000B} ({\tt\string bc03}) for the simple stellar population module assuming a \cite{2003PASP..115..763C} stellar initial mass function. We also use the {\tt\string nebular} module to account for emission from the interstellar medium, as well as the attenuation curve ({\tt\string dustatt\_modified\_starburst}) from \citet{2000ApJ...533..682C}. We follow \citet{2023ApJ...950L...5Y} for the parameter configuration setting, except for the redshift, which is allowed to vary from 0 to 10 in steps of 0.05. 
The best-fit models from {\tt\string EAzY} and {\tt\string CIGALE} yield consistent outcomes (Figure~\ref{off-center_sed_phot-z}).

The derived photometric redshift for MSAID41225 is $z_{\rm phot} \approx 2.7$, with the posterior distribution concentrated between $z_{\rm phot} \approx 2.4$ to 3.6, far from the spectroscopic redshift of the central LRD point source $z_{\rm spec}=6.76$. The off-centered blob near MSAID41225 is very likely an interloping low-redshift galaxy. Given the stellar mass of $\log\, (M_*/M_{\odot})=7.69 \pm 0.24$ derived from {\tt\string CIGALE}, we expect $R_e=0.95\pm 0.66$ kpc from the mass-size relation described in Section~\ref{gs_fitting}, which is consistent with $R_e=0.71^{+0.04}_{-0.03}$ kpc actually measured from our morphological fit. The photometric redshifts of MSAID13821 and MSAID4286 are within 0.5 dex of their spectroscopic redshifts. Even though their redshift posterior distribution functions cover spectroscopic redshifts, they are relatively flat and indicate large uncertainties in the ephotometric redshift estimation. While it is likely that the off-centered emission is associated with the central point source of the LRD, a more definitive conclusion must await stronger evidence.

\begin{figure}[t]
\centering
\includegraphics[width=0.45\textwidth]{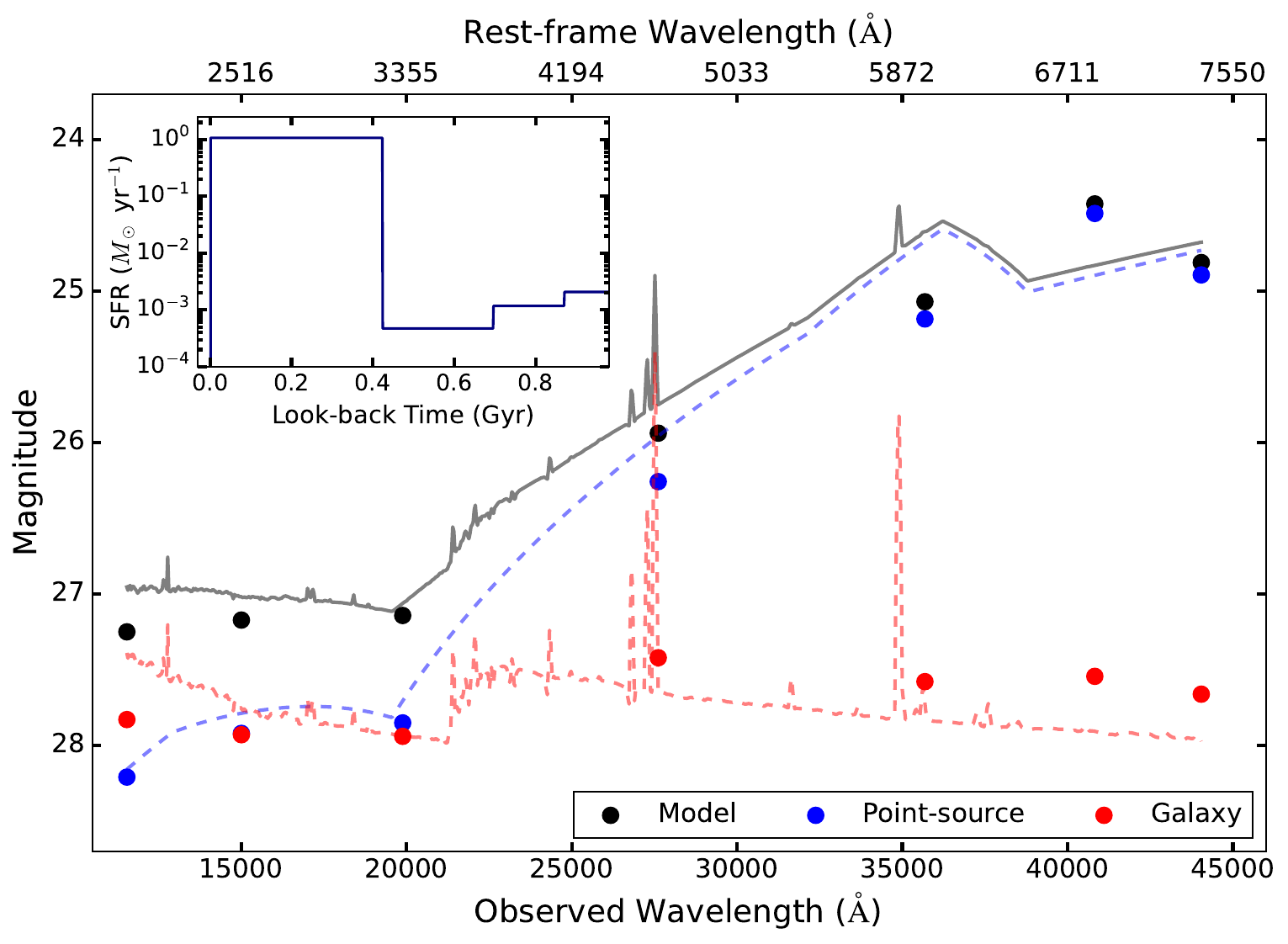}
\caption{SEDs of different components in the best-fit model for MSAID38108, for the entire model (black), central point source (blue), and host galaxy (red). The inset plot presents the star formation history of the host galaxy model.
\label{host_sed_demo}}
\end{figure}

For the other two LRDs, off-centered emission is detected only in the three (MSAID45924) or five (MSAID10686) bluest bands, with the flux generally decreasing with wavelength (Figure~\ref{off-center_sed_measurement}). The non-detection in the redder bands owes to the SED shape and the increase of the PSF size, as well as the rise of the optically red point source, all of which make it harder to disentangle the off-centered emission from the central source. These two blobs have the largest size ($R_e \approx 0.7-1$ kpc) among the five LRDs with off-centered emission. Their profiles are rather flat ($n \approx 0.2-0.7$). Indeed, the shape of the blob near MSAID10686 appears quite irregular (Figure~\ref{off-center_demo}), and a S\'ersic function provides, at best, only a crude approximation. These measurements alone are not enough to determine the nature of the off-centered emission. Nevertheless, we fit their multi-band fluxes and upper limits using {\tt\string CIGALE}. We use the same parameter configurations and model setting as used in other sources, except fixing the redshift to the spectroscopic redshift of the LRDs. In this case, the derived stellar mass of the off-centered emission for MSAID45924 and MSAID10686 are $\log\, (M_*/M_{\odot})=8.72 \pm 0.28$ and $\log\, (M_*/M_{\odot})=7.75 \pm 0.11$, respectively. 

In addition to galaxy SED models, we also explore the possibility that the off-centered emission arises entirely from nebular emission generated by the recombination continuum of diffuse gas. \citet{2024arXiv240907805I} propose that spectral features such as the Balmer break and Balmer absorption observed in some LRDs can be produced by an accretion disk embedded in gas with a extremely high density of $10^{9}-10^{11}\,{\rm cm}^{-3}$. Could the off-centered blobs come from a reservoir of accreted gas near the central engine, or from relic gas associated with the current of past episode of accretion activity? In Figure~\ref{off-center_sed_measurement}, we present fits using pure nebular emission templates generated using {\tt\string CLOUDY} for metalllicity $0.001\leq Z \leq 0.04$ and ionization parameter $-4\leq {\rm log}\, U \leq -2$. {\tt\string CLOUDY} calculations are performed assuming a constant electron density of $100\,{\rm cm}^{-3}$, which is typical for gaseous nebulae \citep{2011MNRAS.415.2920I}. For convenience, we assume that the ionizing radiation comes from the integrated stellar emission spectrum with a constant star formation rate, the value of which is treated as a variable during the fit. Although the actual ionizing radiation for these hypothetical nebulae should be the LRDs themselves, which is very likely dominated by AGN emission, their rest-frame UV radiation can also be approximated using star-forming galaxy models \citep[e.g.,][]{2023arXiv230607320L}. However, as shown in Figure~\ref{off-center_sed_measurement}, while the off-centered emission in MSAID4286 and 13821 can be better described by the galaxy model, neither the galaxy model nor the nebular model can fit the observed SED well in MSAID45924 and 10686. In MSAID45924, nebular templates cannot satisfy the faint flux upper limits observed in the LW filters because of the strong emission lines, while in MSAID10686 nebular model underestimates the flux observed in the SW filters on account of the decrease of the $n=2$ hydrogen recombination continuum toward shorter wavelength. No definitive conclusion can be drawn regarding the nature of these irregularly shaped, off-centered blobs.

\begin{figure}[t]
\centering
\includegraphics[width=0.45\textwidth]{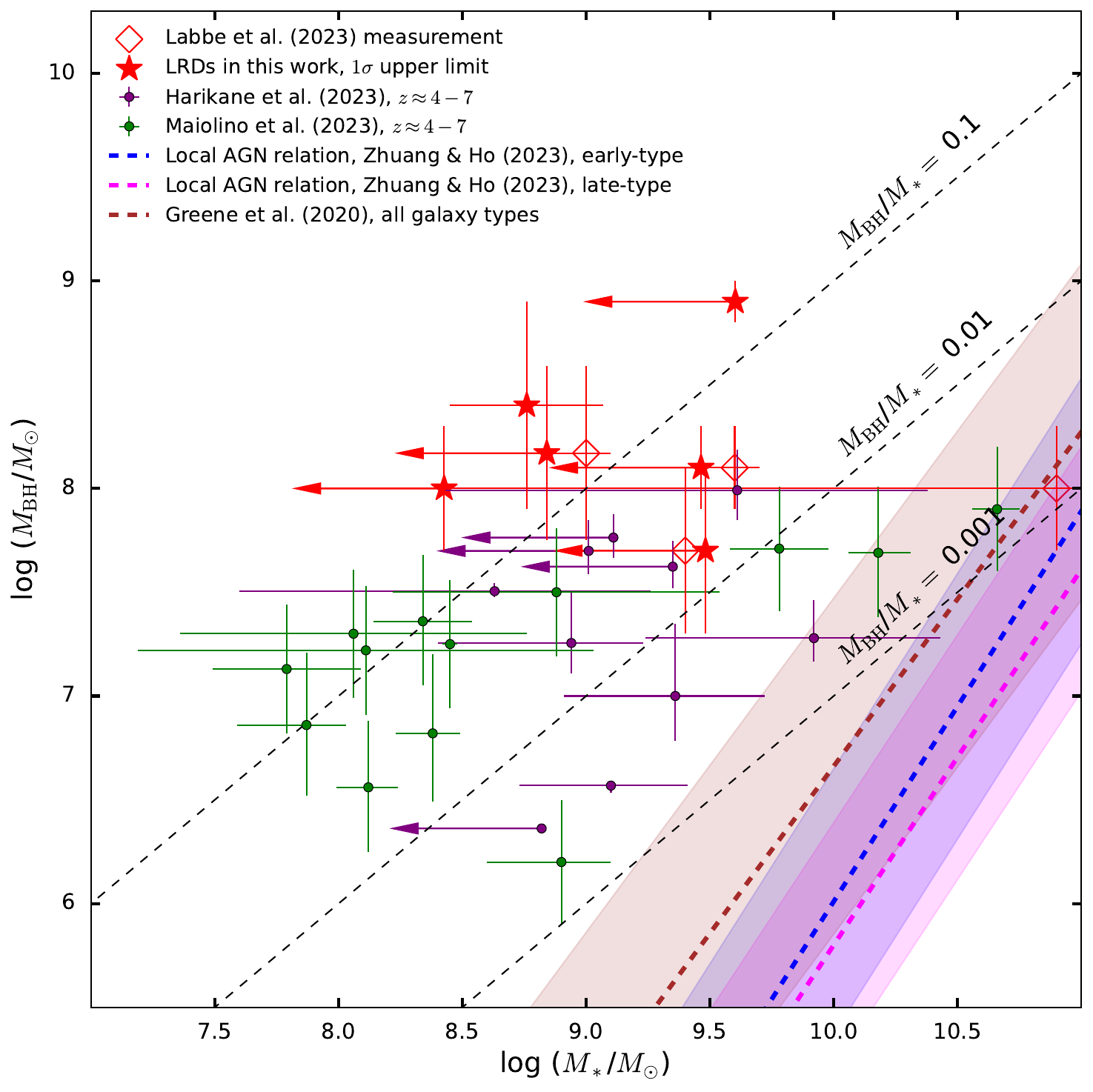}
\caption{The relation between BH mass and the stellar mass of the host galaxy for high-redshift type~1 AGNs. Measurements for the six LRDs with single-epoch BH mass measurements in our sample (red filled stars), along with measurements from \citet[][red open diamonds]{2023arXiv230607320L}. For comparison, we also include the AGN samples studied by \citet[][purple points]{2023ApJ...959...39H} and \citet[][green points]{2023arXiv230801230M}. Colored dashed lines and shaded regions are the local relations and their corresponding uncertainties for local inactive \citep{2020ARA&A..58..257G} and active \citep{2023NatAs...7.1376Z} galaxies. Black dashed lines depict different ratios of BH mass to total stellar mass ($M_{\rm BH}/M_* = 0.001, 0.01$ and 0.1).
\label{mass_scaling_demo}}
\end{figure}

\section{Implications} \label{discussion}

We detect extended emission centered on the point source in only one LRD, MSAID38108 (Figure~\ref{host_detected_demo}). The inferred host galaxy has S\'ersic index $n = 0.71^{+0.07}_{-0.08}$ and effective radius $R_e = 0.66^{+0.08}_{-0.05}$ kpc. The best-fit stellar component yields a stellar mass of $\log\, (M_*/M_{\odot})=8.66^{+0.24}_{-0.23}$ (Figure~\ref{host_sed_demo}). Flux from the host galaxy contributes significantly to the three bands in the SW channel, which correspond to the rest-frame UV portion of the SED, while in the LW channel the rest-frame optical continuum is dominated by the unresolved nucleus. The total continuum from our best-fit model is consistent with the NIRSpec/PRISM spectra in \citet{2024ApJ...964...39G}. Interestingly, the colors calculated using the nuclear SED alone (${\rm F115W}-{\rm F200W}=0.36$, ${\rm F277W}-{\rm F444W}=1.36$, and ${\rm F277W}-{\rm F356W}=1.07$) still satisfy the LRD color selection criteria proposed by \citet{2024ApJ...964...39G} of $(-0.5<{\rm F115W}-{\rm F200W}<1)$ $\wedge$ $({\rm F277W}-{\rm F444W}>1)$ $\wedge$ $({\rm F277W}-{\rm F356W}>0.7)$. No sign of galaxy emission centered on the point source is detected in the remaining LRDs. Two are completely unresolved in all seven NIRCam bands, and the other five show extended emission $\sim 0.4-1$ kpc away from the point source, assuming that they are at the same redshift. In all LRDs, the point source always satisfy the LRD color selection criteria, suggesting that the peculiar ``V-shaped'' SED is an intrinsic property of the central nucleus, despite varying contributions from extended components in the bluer bands. 

These results can be interpreted in different ways, depending on one's assumption regarding the nature of the unresolved central source. The first possibility is that the emission from the point source is dominated by an AGN in all bands. In this case, the lack of extended emission centered on the point source implies that no host galaxy is detected. To explore this scenario further, we estimate a stringent upper limit on the stellar mass of the host using realistic mock simulations, rigorously designed to determine how much stellar mass can be hidden underneath the LRD while remaining unresolved in the rest-frame optical band (Figure~\ref{det_frac_demo}; Table~\ref{results_table}). Figure~\ref{mass_scaling_demo} plots our constraints on the host galaxy stellar mass on the BH mass versus stellar mass diagram, in comparison to the local ($z \approx 0$) scaling relation as well as other high-redshift type~1 AGNs discovered with JWST \citep{2023ApJ...959...39H, 2023arXiv230801230M}. Only the six LRDs with single-epoch BH mass measurements are shown. Assuming that the LRD host galaxies have similar sizes as the general galaxy population at their redshift, our new measurements place the LRDs significantly above the local scaling relations. For example, at a fiducial BH mass of $M_{\rm BH} = 10^8\,M_\odot$, the $M_{\rm BH}-M_*$ relation of \cite{2020ARA&A..58..257G} for all galaxy types predicts $M_{\rm BH}/M_* = 1.5\times 10^{-3}$, whereas for the scaling relation for active galaxies of \cite{2023NatAs...7.1376Z} we expect $M_{\rm BH}/M_* = (0.6-0.9)\times 10^{-3}$, depending on their morphological type. By contrast, our single LRD with a detected host galaxy has $M_{\rm BH}/M_* = 0.55$, a factor of $370-900$ above the local relations; the other seven with no detectable hosts lie $\gtrsim 10-1000$ times above the local relations. Our targets were chosen purposefully to have small lensing magnification, generally with $\mu \approx 1.5$. This translates to maximum enhancement of $\sim 0.2$ dex in stellar mass, which, in any case, would shift the stellar masses even further away from the local scaling relations. 

In terms of the off-centered extended emission discovered around many of the LRDs, even if all of it can be ascribed to the host and originate from stars, the implied stellar mass would not exceed the mass upper limits underneath the point source. The stellar mass estimates would increase by $\lesssim 0.17$ dex, which would hardly affect the overall conclusion that LRDs deviate strongly from the local scaling relation. If the off-centered blobs are, indeed, stellar in origin, they would indicate that the host galaxies of LRDs are not only undermassive relative to their central BH but that the majority of them are strongly lopsided. On the other hand, if the blobs are not part of the LRD hosts, they may be associated with ongoing mergers, as indicated by their sub-kpc separation and irregular morphologies. This would imply a high merger fraction ($50\%$) among the LRD population. 

Another possibility is that a significant fraction of the unresolved flux arises from extremely compact galaxies, such as those discovered by \citet{2023Natur.619..716C} and \citet{2023ApJ...951...72O}, which have typical sizes of $R_e \approx 200-300$ pc, with the most extreme case of $R_e = 39$ pc, consistent with the inferred size limits of LRDs from \citet{2023arXiv230607320L}. In this scenario, the most radical estimate of the stellar mass can be made by assuming that {\it all}\ the emission comes from the host galaxy. This calculation was done by \citet{2023arXiv230607320L} for four of the LRDs in our sample, in which they fit the total SEDs using two stellar population models with different attenuation in order to simultaneously explain the blue UV slope and the red optical slope. Their results are presented alongside ours in Figure~\ref{mass_scaling_demo}. Even in this most extreme case, three of the sources still have $M_{\rm BH}/M_*$ ratios more than an order of magnitude higher than local active and inactive galaxies. The only exception is MSAID4286, which, with $M_{\rm BH} = 10^8\, M_{\odot}$ and $M_* = 10^{10.9}\, M_\odot$ would place it within the locus of local galaxies. However, we note that unlike unlike other sources MSAID4286 does not have an ALMA measurement, nor is its upper limit provided by \citet{2023arXiv230607320L}, which may renders its stellar mass more susceptible to being overestimated by SED fitting.

Our results indicate that LRDs have highly overmassive BHs, or, alternatively but not equivalently, undermassive galaxies. Some caveats to be wary, however. The simulations used to obtain the mass upper limits explicitly assume that the host galaxies hidden underneath the point source follows the mass-size relation of the general high-redshift galaxy population. This might not be the case for AGN host galaxies, if active galaxies are systematically more compact than inactive galaxies of the same mass (e.g., \citealt{2021MNRAS.500.4989N, Zhuang2022}). A more compact galaxy would be harder to detect under the point source. In addition, the mass-to-light ratios for the mock host galaxies are drawn from a high-redshift sample of mostly star-forming galaxies. The galaxy flux contribution to the rest-frame optical would be reduced if the hosts of LRDs are reddened as much as their nuclei. Both of these effects would underestimate the contribution from the host galaxy. 

As a counterbalance, we note that the host galaxy mass upper limits are derived from mock images in a single rest-frame optical band, which usually falls into the LW channel that has lower spatial resolution. As the effective radius of a galaxy is usually larger at shorter wavelengths (e.g., \citealt{2013MNRAS.430..330H}), the the detection probability would be larger and mass upper limits lower than our current measurement had we used a bluer band. With a typical lensing magnification of $\mu \approx 1.5$, the sizes measurements of our sources should not be affected by more than a factor of $\sqrt{\mu} \approx 1.2$. Future work should properly consider the effect of lensing distortion on morphological fitting.

\section{Summary}\label{summary}

We analyze the multi-band morphology and structure of eight LRDs selected in the UNCOVER field. With the aid of {\tt\string GalfitS}, we perform simultaneous multi-band image decomposition while using SED models to constrain the host galaxy flux in different bands. To obtain accurate PSF models, we select field stars whose images are stacked with the same telescope position angle as each LRD observation, to construct hybrid PSF models by combining the central region of an empirical PSF derived from star images with the outskirts of a theoretical PSF.

Our main results are as follows:

\begin{enumerate}

\item Seven out of the eight LRDs are best fit by a single point-source without the need of an additional host galaxy component centered on the nucleus.  In the majority of the LRDs in our sample, we detect no extended emission underneath the LRD point source that can attributed to stellar emission. The only exception, MSAID38108, has detected host emission consistent with a galaxy with a S\'ersic index $n = 0.71^{+0.07}_{-0.08}$, an effective radius $R_e = 0.66^{+0.08}_{-0.05}$ kpc, and a stellar mass $\log\, (M_*/M_{\odot})=8.66^{+0.24}_{-0.23}$.

\item We estimate stringent upper limits for the stellar mass of a hypothetical potential host galaxy by conducting realistic mock simulations that place high-redshift galaxy images under the LRDs. Stellar mass constraints are derived under two extreme scenarios: on the one hand by assuming that all the unresolved emission comes from the AGN, and on the other hand by supposing that all the emission arises from starlight. In either case, the stellar mass limits are at least a factor of 10 lower than expected based on the BH masses estimated from the broad H$\alpha$ emission line and the $z \approx 0$ scaling relation between BH mass and host galaxy stellar mass.

\item We detect extended off-centered emission associated with 50\% of the LRDs. The SEDs of the extended, asymmetric blobs in two sources (MSAID4286 and 13821) can be fit using galaxy templates, while in the other two (MSAID10686 and 45924) neither the galaxy nor pure nebular SED model provides an acceptable fit. The off-centered component, in any case, contributes negligibly to the total stellar mass budget of the host galaxy.

\end{enumerate}

Future works can be improved in several aspects. During the fitting procedure, we adopt the same star formation history across the entire S\'ersic profile, in which case the host galaxy morphology does not vary with wavelength. More sophisticated analysis that considers radial gradients of a parametric star formation history is needed to model realistically the wavelength variation of galaxy morphological. The SED information provided by multi-band images is still quite limited, a problem that can be alleviated by jointly analyzing imaging and spectroscopic data, such as NIRCam slitless or integral-field spectra. The physical nature of the off-centered blobs discovered im half of the LRDs in our sample can be further elucidated if deeper spectroscopic observations targeting on this faint emission can be obtained.

\section*{Acknowledgments}
\begin{acknowledgments}
This work was supported by National Key R\&D Program of China (2022YFF0503401), the National Science Foundation of China (11721303, 11991052, 12011540375, 12233001) and the China Manned Space Project (CMS-CSST-2021-A04, CMS-CSST-2021-A06). We thank Kohei Inayoshi and the referee for helpful comments.
\end{acknowledgments}

\appendix

\section{Fitting Results}\label{app_a}

We present the fitting results for the remaining five LRDs in our sample. The fitting follows the procedures described in Section~\ref{gs_fitting}. The results are illusttated in Figures~A1--A5, which follow the conventions and format of Figure~\ref{2008_fitting_demo}.

\begin{figure*}[b!]
\figurenum{A1}
\centering
\includegraphics[width=0.75\textwidth]{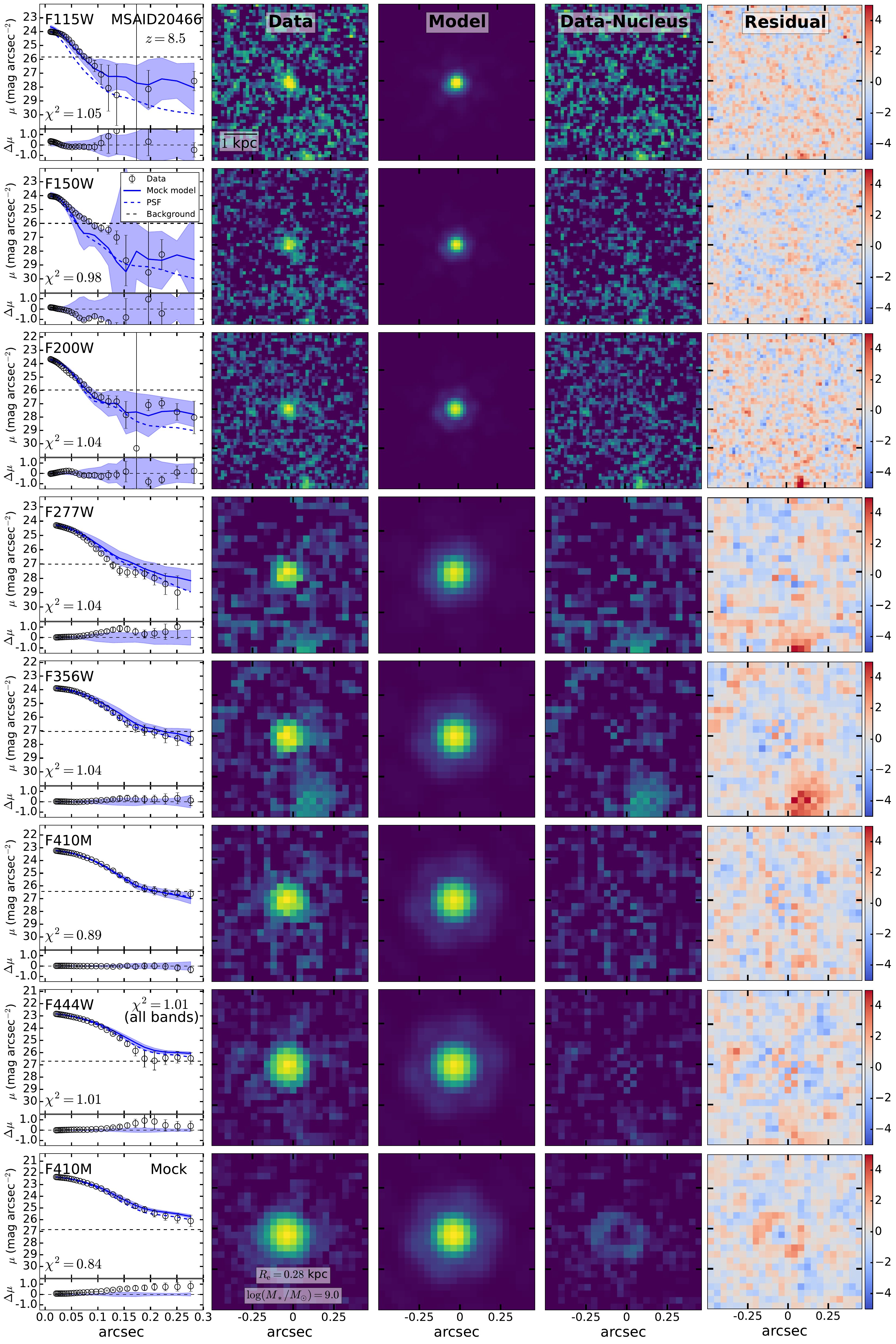}
\caption{Same as in Figure~\ref{2008_fitting_demo}.
\label{20466_fitting_demo}}
\end{figure*}

\begin{figure*}[b!]
\figurenum{A2}
\centering
\includegraphics[width=0.75\textwidth]{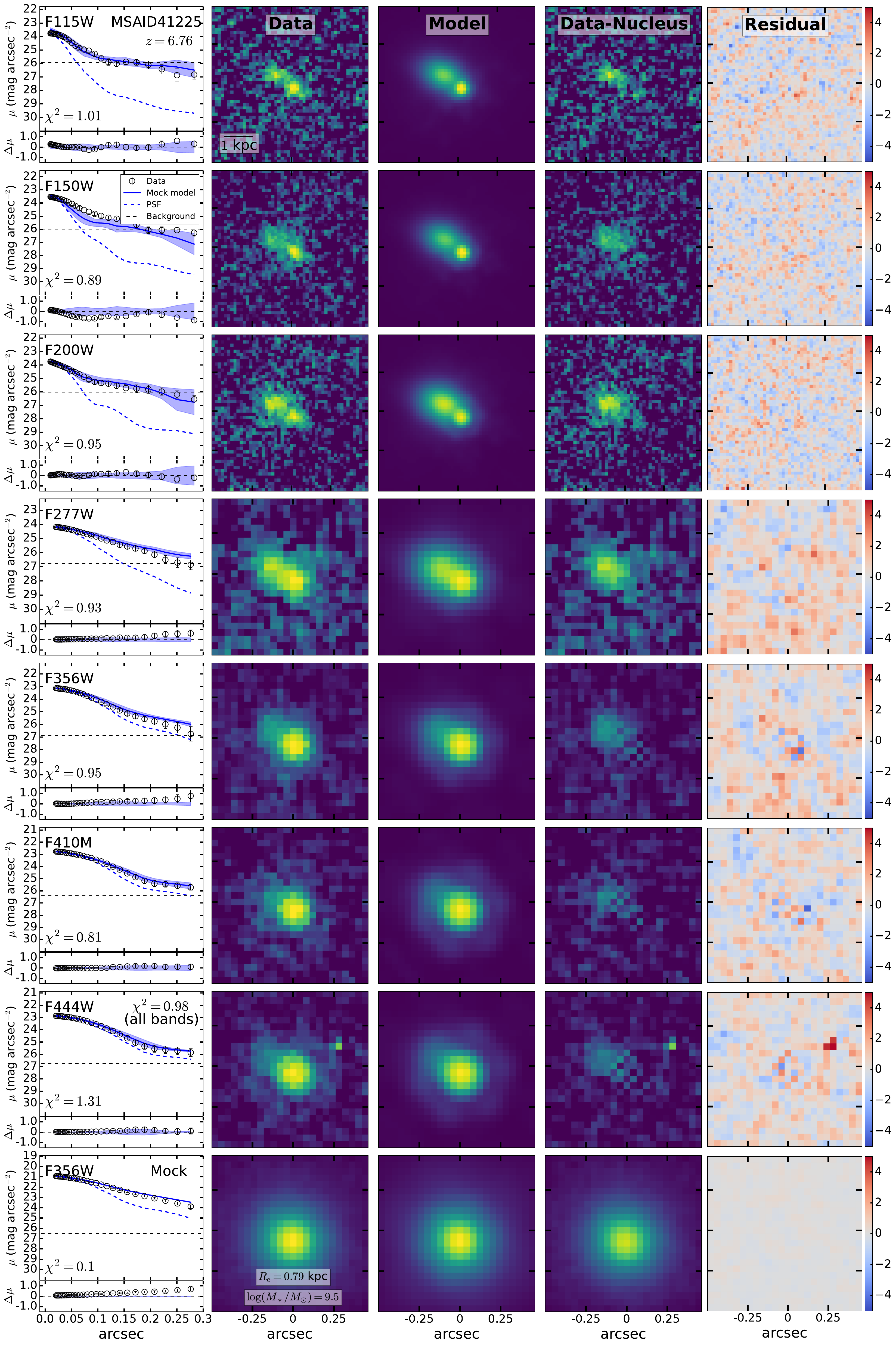}
\caption{Same as in Figure~\ref{2008_fitting_demo}.
\label{41225_fitting_demo}}
\end{figure*}

\begin{figure*}[b!]
\figurenum{A3}
\centering
\includegraphics[width=0.75\textwidth]{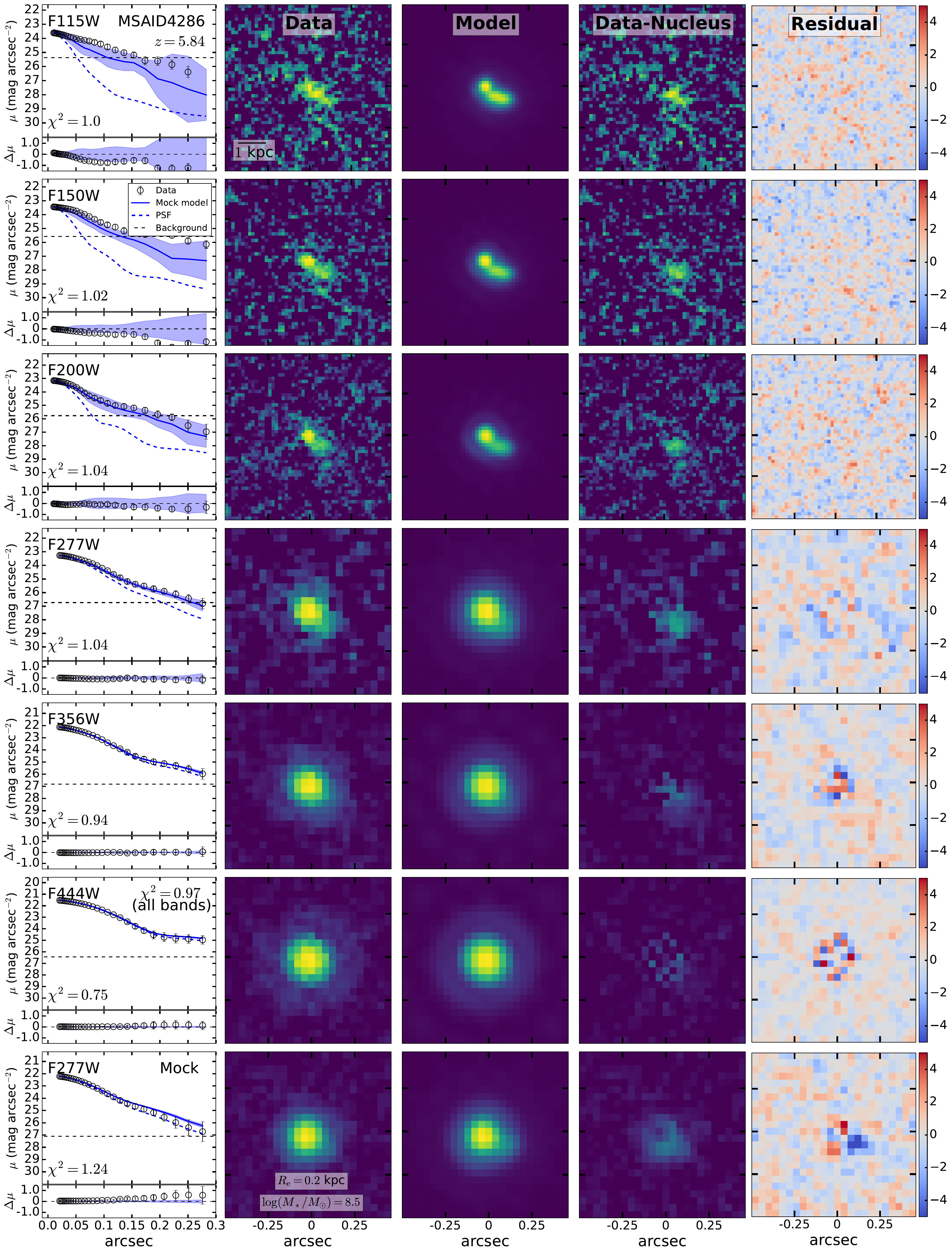}
\caption{Same as in Figure~\ref{2008_fitting_demo}.
\label{4286_fitting_demo}}
\end{figure*}

\begin{figure*}[b!]
\figurenum{A4}
\centering
\includegraphics[width=0.75\textwidth]{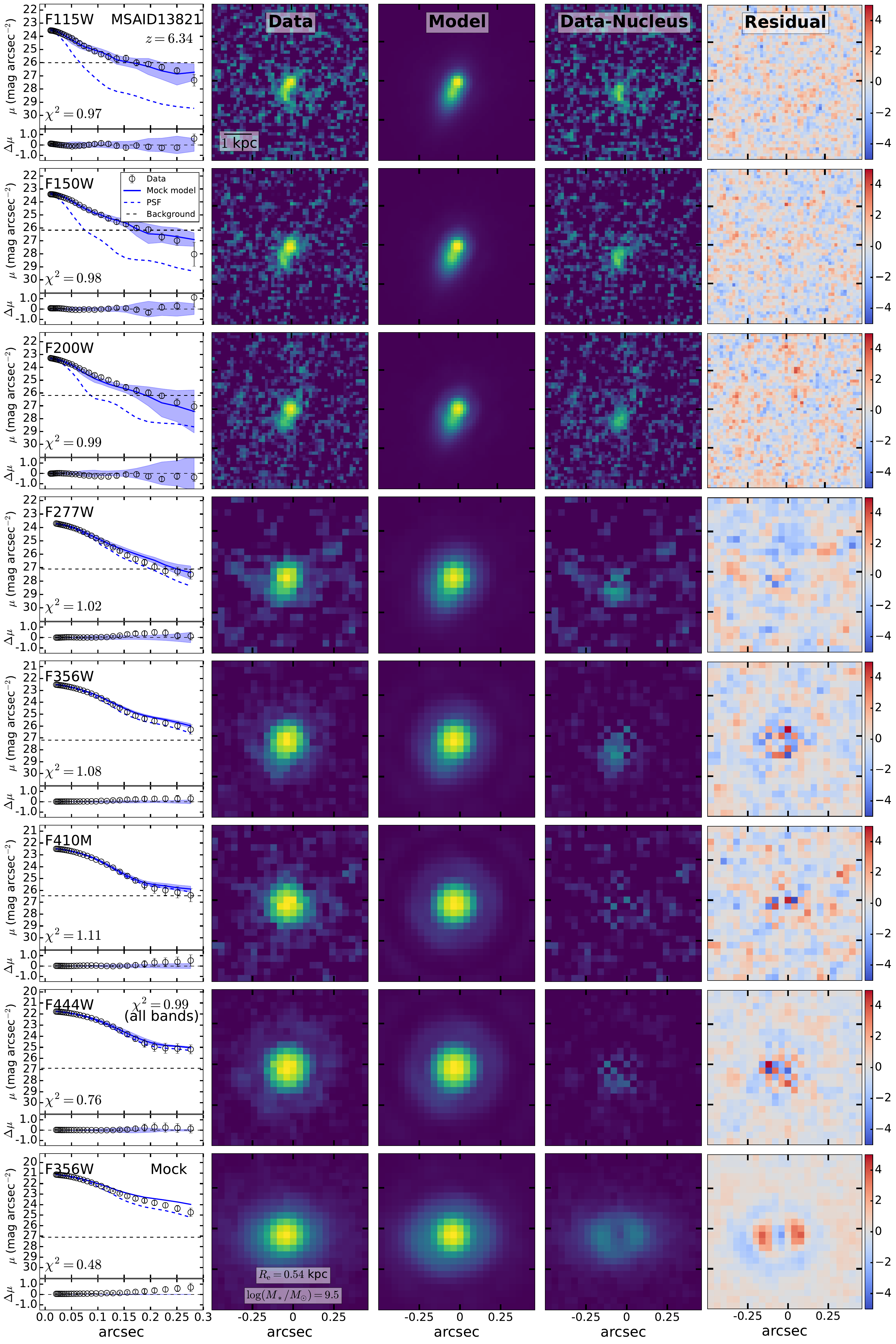}
\caption{Same as in Figure~\ref{2008_fitting_demo}.
\label{13821_fitting_demo}}
\end{figure*}

\begin{figure*}[b!]
\figurenum{A5}
\centering
\includegraphics[width=0.75\textwidth]{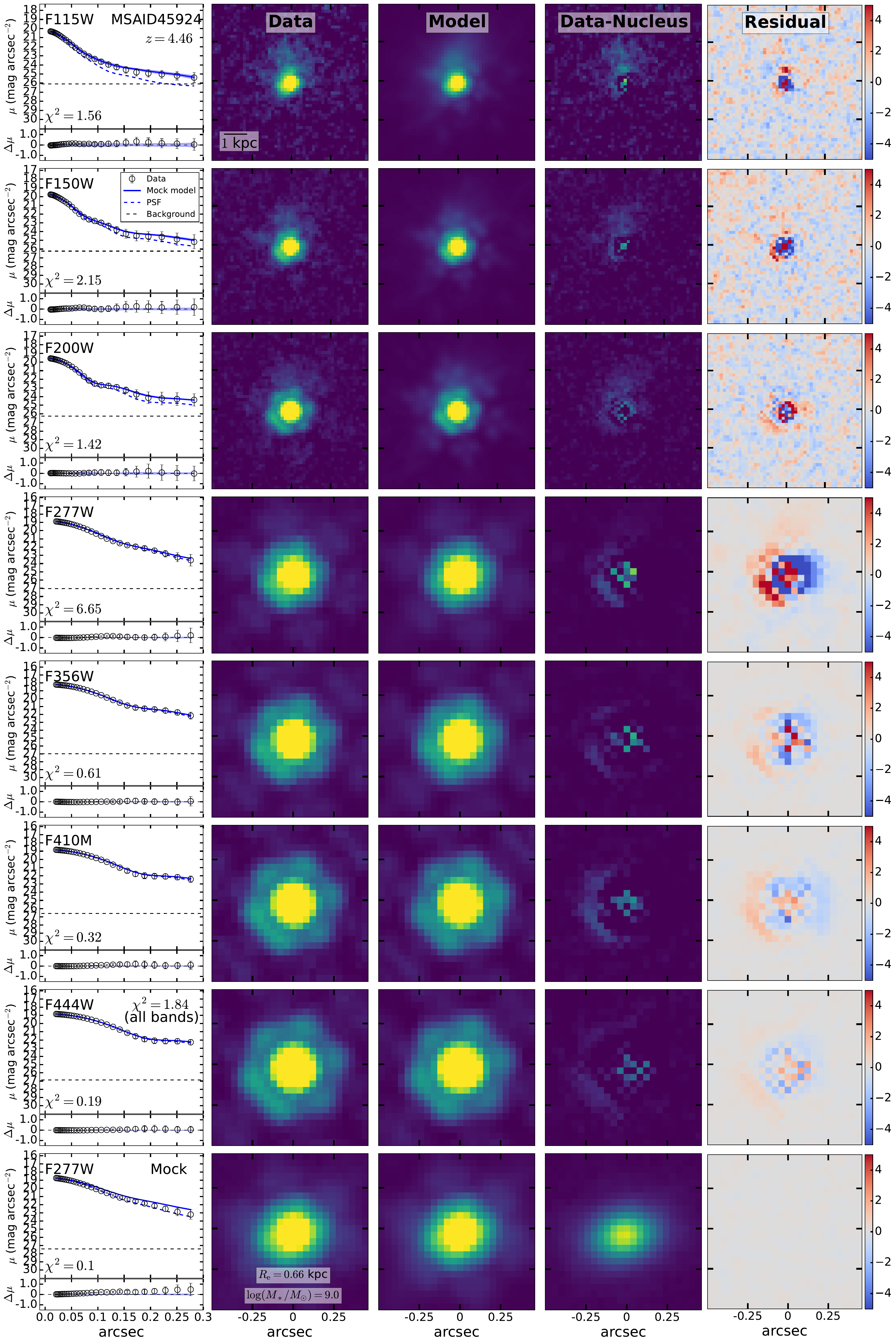}
\caption{Same as in Figure~\ref{2008_fitting_demo}.
\label{45924_fitting_demo}}
\end{figure*}

\clearpage

\newpage


\end{CJK*}
\end{document}